\documentclass[journal, web]{ieee_tcns_template}
\usepackage{ieee_tcns_style}
\usepackage[utf8]{inputenc} 
\usepackage[T1]{fontenc}    
\usepackage{times}
\usepackage{soul}
\usepackage{cite}
{
\newtheorem{assumption}{Assumption}
\newtheorem{proposition}{Proposition}
}
\usepackage{multirow}
\usepackage[linesnumbered,ruled,vlined]{algorithm2e}
\usepackage{algpseudocode}
\algrenewcommand\algorithmicrequire{\textbf{Input:}}
\algrenewcommand\algorithmicensure{\textbf{Output:}}
\SetKwRepeat{Do}{do}{while}
\usepackage{threeparttable}
\usepackage{makecell}
\usepackage{amsmath,amssymb,amsfonts,dsfont}
\usepackage{graphicx}
\usepackage[singlelinecheck=false,justification=justified]{caption}
\usepackage[outdir=./]{epstopdf}
\usepackage{textcomp}
\usepackage{xcolor}

\usepackage{mathtools}

\usepackage{subfigure}
\renewcommand{\theenumi}{\alph{enumi}}
\SetKwInput{KwInput}{Input}                
\SetKwInput{KwOutput}{Output}              
\def\BibTeX{{\rm B\kern-.05em{\sc i\kern-.025em b}\kern-.08em
    T\kern-.1667em\lower.7ex\hbox{E}\kern-.125emX}}

\usepackage{booktabs}       
\usepackage{nicefrac}       
\usepackage{microtype}      
\usepackage{wrapfig}
\usepackage[breaklinks=true, colorlinks, bookmarks=true]{hyperref}

\newtheorem{theorem}{Theorem}
\newtheorem{lemma}{Lemma}

\DeclareMathOperator*{\diag}{diag}
\DeclareMathOperator*{\dist}{dist}
\DeclareMathOperator*{\rank}{rank}
\newcommand{\scaleMathLine}[2][1]{\resizebox{#1\linewidth}{!}{$\displaystyle{#2}$}}

\newcommand{\revision}[1]{{\color{black}#1}}

\begin{document}

\title{Distributed Multi-Agent Reinforcement Learning with One-hop Neighbors and Compute Straggler Mitigation}

\author{Baoqian~Wang,~\IEEEmembership{Student Member,~IEEE}, 
    Junfei~Xie,~\IEEEmembership{Senior Member,~IEEE}, 
    Nikolay Atanasov,~\IEEEmembership{Senior Member,~IEEE}%
\thanks{We gratefully acknowledge support from NSF CAREER-2048266, NSF CCF-2402689 and ONR N00014-23-1-2353.}%
\thanks{Baoqian Wang is with the Boeing AI, The Boeing Company, Huntsville, AL 35808, USA (e-mail: {\tt\small baoqian.wang@boeing.com)}.}%
\thanks{Junfei Xie is with the Department of Electrical and Computer Engineering, San Diego State University, San Diego, CA 92182, USA (e-mail: {\tt\small jxie4@sdsu.edu}). Corresponding author.}%
\thanks{Nikolay Atanasov is with the Department of Electrical and Computer Engineering, University of California San Diego, La Jolla, CA 92093, USA (e-mail: \tt\small natanasov@ucsd.edu).}}

\maketitle
\newcommand{\calA}{{\cal A}}
\newcommand{\calB}{{\cal B}}
\newcommand{\calC}{{\cal C}}
\newcommand{\calD}{{\cal D}}
\newcommand{\calE}{{\cal E}}
\newcommand{\calF}{{\cal F}}
\newcommand{\calG}{{\cal G}}
\newcommand{\calH}{{\cal H}}
\newcommand{\calI}{{\cal I}}
\newcommand{\calJ}{{\cal J}}
\newcommand{\calK}{{\cal K}}
\newcommand{\calL}{{\cal L}}
\newcommand{\calM}{{\cal M}}
\newcommand{\calN}{{\cal N}}
\newcommand{\calO}{{\cal O}}
\newcommand{\calP}{{\cal P}}
\newcommand{\calQ}{{\cal Q}}
\newcommand{\calR}{{\cal R}}
\newcommand{\calS}{{\cal S}}
\newcommand{\calT}{{\cal T}}
\newcommand{\calU}{{\cal U}}
\newcommand{\calV}{{\cal V}}
\newcommand{\calW}{{\cal W}}
\newcommand{\calX}{{\cal X}}
\newcommand{\calY}{{\cal Y}}
\newcommand{\calZ}{{\cal Z}}

\newcommand{\setA}{\textsf{A}}
\newcommand{\setB}{\textsf{B}}
\newcommand{\setC}{\textsf{C}}
\newcommand{\setD}{\textsf{D}}
\newcommand{\setE}{\textsf{E}}
\newcommand{\setF}{\textsf{F}}
\newcommand{\setG}{\textsf{G}}
\newcommand{\setH}{\textsf{H}}
\newcommand{\setI}{\textsf{I}}
\newcommand{\setJ}{\textsf{J}}
\newcommand{\setK}{\textsf{K}}
\newcommand{\setL}{\textsf{L}}
\newcommand{\setM}{\textsf{M}}
\newcommand{\setN}{\textsf{N}}
\newcommand{\setO}{\textsf{O}}
\newcommand{\setP}{\textsf{P}}
\newcommand{\setQ}{\textsf{Q}}
\newcommand{\setR}{\textsf{R}}
\newcommand{\setS}{\textsf{S}}
\newcommand{\setT}{\textsf{T}}
\newcommand{\setU}{\textsf{U}}
\newcommand{\setV}{\textsf{V}}
\newcommand{\setW}{\textsf{W}}
\newcommand{\setX}{\textsf{X}}
\newcommand{\setY}{\textsf{Y}}
\newcommand{\setZ}{\textsf{Z}}

\newcommand{\bfa}{\mathbf{a}}
\newcommand{\bfb}{\mathbf{b}}
\newcommand{\bfc}{\mathbf{c}}
\newcommand{\bfd}{\mathbf{d}}
\newcommand{\bfe}{\mathbf{e}}
\newcommand{\bff}{\mathbf{f}}
\newcommand{\bfg}{\mathbf{g}}
\newcommand{\bfh}{\mathbf{h}}
\newcommand{\bfi}{\mathbf{i}}
\newcommand{\bfj}{\mathbf{j}}
\newcommand{\bfk}{\mathbf{k}}
\newcommand{\bfl}{\mathbf{l}}
\newcommand{\bfm}{\mathbf{m}}
\newcommand{\bfn}{\mathbf{n}}
\newcommand{\bfo}{\mathbf{o}}
\newcommand{\bfp}{\mathbf{p}}
\newcommand{\bfq}{\mathbf{q}}
\newcommand{\bfr}{\mathbf{r}}
\newcommand{\bfs}{\mathbf{s}}
\newcommand{\bft}{\mathbf{t}}
\newcommand{\bfu}{\mathbf{u}}
\newcommand{\bfv}{\mathbf{v}}
\newcommand{\bfw}{\mathbf{w}}
\newcommand{\bfx}{\mathbf{x}}
\newcommand{\bfy}{\mathbf{y}}
\newcommand{\bfz}{\mathbf{z}}

\newcommand{\bfalpha}{\boldsymbol{\alpha}}
\newcommand{\bfbeta}{\boldsymbol{\beta}}
\newcommand{\bfgamma}{\boldsymbol{\gamma}}
\newcommand{\bfdelta}{\boldsymbol{\delta}}
\newcommand{\bfepsilon}{\boldsymbol{\epsilon}}
\newcommand{\bfzeta}{\boldsymbol{\zeta}}
\newcommand{\bfeta}{\boldsymbol{\eta}}
\newcommand{\bftheta}{\boldsymbol{\theta}}
\newcommand{\bfiota}{\boldsymbol{\iota}}
\newcommand{\bfkappa}{\boldsymbol{\kappa}}
\newcommand{\bflambda}{\boldsymbol{\lambda}}
\newcommand{\bfmu}{\boldsymbol{\mu}}
\newcommand{\bfnu}{\boldsymbol{\nu}}
\newcommand{\bfomicron}{\boldsymbol{\omicron}}
\newcommand{\bfpi}{\boldsymbol{\pi}}
\newcommand{\bfrho}{\boldsymbol{\rho}}
\newcommand{\bfsigma}{\boldsymbol{\sigma}}
\newcommand{\bftau}{\boldsymbol{\tau}}
\newcommand{\bfupsilon}{\boldsymbol{\upsilon}}
\newcommand{\bfphi}{\boldsymbol{\phi}}
\newcommand{\bfchi}{\boldsymbol{\chi}}
\newcommand{\bfpsi}{\boldsymbol{\psi}}
\newcommand{\bfomega}{\boldsymbol{\omega}}
\newcommand{\bfxi}{\boldsymbol{\xi}}
\newcommand{\bfell}{\boldsymbol{\ell}}

\newcommand{\bfA}{\mathbf{A}}
\newcommand{\bfB}{\mathbf{B}}
\newcommand{\bfC}{\mathbf{C}}
\newcommand{\bfD}{\mathbf{D}}
\newcommand{\bfE}{\mathbf{E}}
\newcommand{\bfF}{\mathbf{F}}
\newcommand{\bfG}{\mathbf{G}}
\newcommand{\bfH}{\mathbf{H}}
\newcommand{\bfI}{\mathbf{I}}
\newcommand{\bfJ}{\mathbf{J}}
\newcommand{\bfK}{\mathbf{K}}
\newcommand{\bfL}{\mathbf{L}}
\newcommand{\bfM}{\mathbf{M}}
\newcommand{\bfN}{\mathbf{N}}
\newcommand{\bfO}{\mathbf{O}}
\newcommand{\bfP}{\mathbf{P}}
\newcommand{\bfQ}{\mathbf{Q}}
\newcommand{\bfR}{\mathbf{R}}
\newcommand{\bfS}{\mathbf{S}}
\newcommand{\bfT}{\mathbf{T}}
\newcommand{\bfU}{\mathbf{U}}
\newcommand{\bfV}{\mathbf{V}}
\newcommand{\bfW}{\mathbf{W}}
\newcommand{\bfX}{\mathbf{X}}
\newcommand{\bfY}{\mathbf{Y}}
\newcommand{\bfZ}{\mathbf{Z}}

\newcommand{\bfGamma}{\boldsymbol{\Gamma}}
\newcommand{\bfDelta}{\boldsymbol{\Delta}}
\newcommand{\bfTheta}{\boldsymbol{\Theta}}
\newcommand{\bfLambda}{\boldsymbol{\Lambda}}
\newcommand{\bfPi}{\boldsymbol{\Pi}}
\newcommand{\bfSigma}{\boldsymbol{\Sigma}}
\newcommand{\bfUpsilon}{\boldsymbol{\Upsilon}}
\newcommand{\bfPhi}{\boldsymbol{\Phi}}
\newcommand{\bfPsi}{\boldsymbol{\Psi}}
\newcommand{\bfOmega}{\boldsymbol{\Omega}}

\newcommand{\bbA}{\mathbb{A}}
\newcommand{\bbB}{\mathbb{B}}
\newcommand{\bbC}{\mathbb{C}}
\newcommand{\bbD}{\mathbb{D}}
\newcommand{\bbE}{\mathbb{E}}
\newcommand{\bbF}{\mathbb{F}}
\newcommand{\bbG}{\mathbb{G}}
\newcommand{\bbH}{\mathbb{H}}
\newcommand{\bbI}{\mathbb{I}}
\newcommand{\bbJ}{\mathbb{J}}
\newcommand{\bbK}{\mathbb{K}}
\newcommand{\bbL}{\mathbb{L}}
\newcommand{\bbM}{\mathbb{M}}
\newcommand{\bbN}{\mathbb{N}}
\newcommand{\bbO}{\mathbb{O}}
\newcommand{\bbP}{\mathbb{P}}
\newcommand{\bbQ}{\mathbb{Q}}
\newcommand{\bbR}{\mathbb{R}}
\newcommand{\bbS}{\mathbb{S}}
\newcommand{\bbT}{\mathbb{T}}
\newcommand{\bbU}{\mathbb{U}}
\newcommand{\bbV}{\mathbb{V}}
\newcommand{\bbW}{\mathbb{W}}
\newcommand{\bbX}{\mathbb{X}}
\newcommand{\bbY}{\mathbb{Y}}
\newcommand{\bbZ}{\mathbb{Z}}
\newcommand{\Var}{\mathrm{Var}}

\begin{abstract}
Most multi-agent reinforcement learning (MARL) methods are limited in the scale of problems they can handle. With increasing numbers of agents, the number of training iterations required to find the optimal behaviors increases exponentially due to the exponentially growing joint state and action spaces. This paper tackles this limitation by introducing a scalable MARL method called Distributed multi-Agent Reinforcement Learning with One-hop Neighbors (DARL1N). DARL1N is an off-policy actor-critic method that addresses the curse of dimensionality by restricting information exchanges among the agents to one-hop neighbors when representing value and policy functions. Each agent optimizes its value and policy functions over a one-hop neighborhood, significantly reducing the learning complexity, yet maintaining expressiveness by training with varying neighbor numbers and states. This structure allows us to formulate a distributed learning framework to further speed up the training procedure. Distributed computing systems, however, contain \emph{straggler} compute nodes, which are slow or unresponsive due to communication bottlenecks, software or hardware problems. To mitigate the detrimental straggler effect, we introduce a novel coded distributed learning architecture,
which leverages coding theory to improve the resilience of the learning system to stragglers. 
Comprehensive experiments show that DARL1N significantly reduces training time without sacrificing policy quality and is scalable as the number of agents increases. Moreover, the coded distributed learning architecture improves training efficiency in the presence of stragglers. 
\end{abstract}

\begin{IEEEkeywords}
Multi-Agent Reinforcement Learning, Scalability, Distributed Computing,  Coded Computation.
\end{IEEEkeywords}
\section{Introduction}
Recent years have witnessed tremendous success of reinforcement learning (RL) in challenging decision making problems, such as robot control and video games. 
Research efforts are currently focused on multi-agent settings, including cooperative robot navigation \cite{jin2021hierarchical}, multi-player games \cite{song2016off}, and traffic management \cite{antonio2022multi}. Direct application of RL techniques in multi-agent settings by running single-agent algorithms simultaneously on each agent exhibits poor performance \cite{lowe2017multi}. This is because, without considering interactions among the agents, the environment becomes non-stationary from the perspective of a single agent.

Multi-agent reinforcement learning (MARL) \cite{bucsoniu2010multi} addresses this challenge by considering all agents and their dynamics collectively when learning the value function and policy of an individual agent. Most effective MARL algorithms, such as multi-agent deep deterministic policy gradient (MADDPG) \cite{lowe2017multi} and counterfactual multi-agent (COMA) \cite{foerster2018counterfactual}, adopt this strategy. However, learning a joint-state value or action-value (Q) or policy function is challenging due to the exponentially growing joint state and action spaces with increasing number of agents \cite{gogineni2023scalability, qu2020scalable}. Policies trained with joint state-action pairs have poor performance in large-scale settings as demonstrated in recent work \cite{yang2018mean,long2020evolutionary}, because their accurate approximation requires models with extremely large capacity.

MARL algorithms that improve the quality of learned policies for large-scale multi-agent settings often employ value function factorization that factorizes the global value/Q function into individual value/Q functions depending on local states and actions, e.g., as in Value Decomposition Network (VDN) \cite{sunehag2017value}, QMIX \cite{rashid2018qmix}, QTran \cite{son2019qtran}, mean-field MARL (MFAC) \cite{yang2018mean} or scalable actor critic (SAC) \cite{qu2020scalable}. In addition to value factorization, there are several other methods proposed to enable scalable MARL. MAAC \cite{iqbal2019actor} uses an attention module to abstract states of other agents when training an agent's Q function, which reduces the quadratically increasing input space to a linear space. EPC \cite{long2020evolutionary} applies curriculum learning to gradually scale MARL up. While these methods achieve great policy performance, the training time can be significant when the number of agents increases because these methods cannot be easily trained in an efficient distributed or parallel manner over multiple computers. 

To address the challenge of training policies for large numbers of agents over a distributed computing architecture, we propose a MARL algorithm called Distributed multi-Agent Reinforcement Learning with One-hop Neighbors (DARL1N). DARL1N's \emph{main advantage} over state-of-the-art MARL methods is that it allows distributed training \revision{across compute nodes (devices with networking, storage, and computing capabilities) running in parallel}, with each compute node simulating only a very small subset of the agent transitions. This is made possible by modeling the agent team topology as a proximity graph and representing the Q function and policy of each agent as a function of its one-hop neighbors only. This structure significantly reduces the representation complexity of the Q and policy functions and yet maintains expressiveness when training is done over varying states and numbers of neighbors. Furthermore, when agent interactions are restricted to one-hop neighborhoods, training an agent's Q function and policy requires transitions only of the agent itself and its potential two-hop neighbors. This enables highly efficient distributed training because each compute node needs to simulate only the transitions of the agents assigned to it and their two-hop neighbors.

\revision{RL or MARL policies can be trained over a distributed computing architecture in either a centralized \cite{nair2015massively, wang2021coding} or decentralized \cite{zhang2018fully} manner. 
Decentralized architectures \cite{zhang2018fully} offer greater resilience to node failures and malicious attacks but 
introduce significant communication overhead as frequent information exchanges are required. Additionally, achieving global coordination in such systems is inherently challenging: global information must either be inferred under the assumption of globally observable states, which limits its applicability, or obtained through consensus, which is difficult to achieve in large-scale systems. 
In contrast, centralized architectures with a central controller are more communication-efficient and facilitate easier global coordination, with  
communication 
occurring 
either asynchronously \cite{mnih2016asynchronous} or synchronously \cite{a2c}.} Asynchronous training faces multiple challenges including slow convergence, difficult debugging and analysis, and sometimes subpar quality of learned policies as learners may return stale gradients evaluated with old parameters \cite{ho2013more,li2014communication,dutta2018slow,a2c,tan1993multi}. \revision{Synchronous training is superior in these aspects but is vulnerable to computing \emph{stragglers} \cite{reisizadeh2019coded} that are common in wireless and mobile networks. These stragglers, which are slow or unresponsive compute nodes caused by communication bottlenecks or software and hardware issues, can result in delays or failures in the training process.} Coded computation \cite{lee18speeding} that employs coding theory to introduce redundant computation can mitigate computing stragglers. While extensively explored in various distributed computation problems such as matrix multiplication \cite{lee18speeding}, linear inverse problems \cite{yang2017coded}, convolution \cite{zhou2022dynamic}, and map reduce \cite{li2015coded}, its application for MARL remains under-studied. In our previous work \cite{wang2021coding}, we explored the merits of coded computation in enhancing resilience and accelerating the training of MADDPG \cite{wang2021coding} in a distributed manner. Building upon this, in this paper, we propose a coded distributed learning architecture tailored for DARL1N. Unlike the one introduced in  \cite{wang2021coding}, where the central controller simulates global environment interactions among all agents and sends simulation data to each learner to train an agent, in the new architecture, each learner directly simulates local environment interactions among a small set of neighboring agents during individual agent training and thus improves distributed computing efficiency and reduces communication overhead.

\textbf{Contributions:}
The primary contribution of this paper is a new MARL algorithm called DARL1N, which employs one-hop neighborhood factorization of the value and policy functions, allowing distributed training with each compute node simulating a small number of agent transitions. DARL1N supports \emph{highly-efficient distributed training} and generates \emph{high-quality multi-agent policies for large agent teams}. The second contribution is a novel coded distributed learning architecture for DARL1N called Coded DARL1N, which allows individual agents to be trained by multiple compute nodes simultaneously, enabling \emph{resilience to stragglers}. Our analysis shows that introducing redundant computations via coding theory does not introduce bias in the value and policy gradient estimates, and the training converges similarly to stochastic gradient descent-based methods. Four codes including Maximum Distance Separable (MDS), Random Sparse, Repetition, and Low Density Generator Matrix (LDGM) codes are investigated to introduce redundant computation. Moreover, we conduct comprehensive experiments comparing DARL1N with four state-of-the-art MARL methods, including MADDPG, MFAC, EPC and SAC, and evaluating their performance in different RL environments, including Ising Model, Food Collection, Grassland, Adversarial Battle, and Multi-Access Wireless Communication. We also conduct experiments to evaluate the resilience of Coded DARL1N to stragglers when trained under different coding schemes. 

It is noted that DARL1N  was first presented in a short conference version \cite{wang2022darl1n}. 
Differing from this version, this journal article 
further extends DARL1N by introducing a new coded distributed learning architecture to enhance its resilience to stragglers while also improving training efficiency. Theoretical analysis is conducted to elucidate the convergence of Coded DARL1N. 
Additionally, this journal undertakes a more comprehensive experimental study, encompassing not only the performance of DARL1N but also its new coded variant. It includes additional benchmarks, environments, and evaluation metrics for a more thorough assessment from various aspects. 


In the rest of the paper, Sec. \ref{sec:problem} formulates the MARL problem to be addressed and describes the occurrence of stragglers within distributed learning systems. The proposed DARL1N algorithm is then introduced in Sec. \ref{sec:method}, followed by the coded distributed learning architecture and different coding schemes, which are described in Sec. \ref{sec:training_architecture} and Sec. \ref{sec:assignment}, respectively. Experiment results are presented in Sec. \ref{sec:experiments}. Limitations and future work are discussed in Sec. \ref{sec::limitation}. Finally, Sec. \ref{sec:conclusion} concludes the paper. 
\section{Problem Statement}
\label{sec:problem}

In MARL, $M$ agents learn to optimize their behavior by interacting with the environment. Denote the state and action of agent $i \in [M] := \{1,\ldots,M\}$  by $s_{i}\in \mathcal{S}_i$ and $a_{i} \in \calA_i$, respectively, where $\calS_i$ and $\calA_i$ are the corresponding state and action spaces. Let $\bfs := (s_{1},\ldots, s_{M}) \in \mathcal{S} := \prod_{i \in [M]} \mathcal{S}_{i}$ and $\bfa := (a_{1}, \ldots, a_{M}) \in \mathcal{A} := \prod_{i \in [M]} \mathcal{A}_{i}$ denote the joint state and action of all agents. At time $t$, a joint action $\bfa(t)$ applied at state $\bfs(t)$ triggers a transition to a new state $\bfs(t+1) \in \calS$ according to a conditional probability density function (pdf) $p(\bfs(t+1) | \bfs(t), \bfa(t))$. After each transition, each agent $i$ receives a reward $r_i(\bfs(t),\bfa(t))$, determined by the joint state and action according to the function $r_i : \calS \times \calA \mapsto \bbR$. The objective of each agent $i$ is to design a policy $\mu_i:\calS \rightarrow \calA_i$ to maximize the expected cumulative discounted reward (known as the \emph{value function}):
\begin{equation}
\label{eq:value_function}
V^{\bfmu}_i(\bfs) := \mathbb{E}_{\substack{ \bfa(t)= \bfmu(\bfs(t))\\\bfs(t)\sim p\phantom{(\bfs(t))}}} \biggl[\sum_{t=0}^{\infty}\gamma^t r_i(\bfs(t),\bfa(t)) \;\big\vert\; \bfs(0)=\bfs\biggr],\nonumber
\end{equation}
where $\bfmu := \left(\mu_{1}, \ldots, \mu_{M}\right)$ denotes the joint policy of all agents and $\gamma \in (0,1)$ is a discount factor. Alternatively, an optimal policy $\mu_i^*$ for agent $i$ can be obtained by maximizing the \emph{action-value (Q) function}:
\begin{align}
\label{eq:qvalue_function}
 Q_i^{\bfmu}&(\bfs,\bfa):=\nonumber\\&  \mathbb{E}_{\substack{ \bfa(t)= \bfmu(\bfs(t))\\\bfs(t)\sim p\phantom{(\bfs(t))}}}\biggl[\sum_{t=0}^{\infty}\gamma^t r_i(\bfs(t),\bfa(t)) \;\big\vert\; \bfs(0)=\bfs, \bfa(0) = \bfa\biggr]\nonumber
\end{align}
and setting $\mu_i^*(\bfs) \in  \arg\max_{a_i} \max_{\bfa_{-i}} Q_i^{*}(\bfs,\bfa)$, where $Q_i^{*}(\bfs,\bfa) := \max_{\bfmu} Q_i^{\bfmu}(\bfs,\bfa)$ and $\bfa_{-i}$ denotes the actions of all agents except $i$. 


To develop a distributed MARL algorithm, we impose additional structure on the MARL problem. Assume that all agents share a common state space, i.e., $\calS_i = \calS_j$, $\forall i,j \in [M]$ and let $\dist: \calS_i\times\calS_i\rightarrow\mathbb{R}$ be a distance metric on the state space. \revision{Note that the distance metric can also be defined over a common state subspace. 
However, for notation simplicity, a common state space is assumed here.} 
Consider a proximity graph \cite{bullo2009distributed} that models the topology of the agent team. A $d$-disk proximity graph is defined as a mapping that associates the joint state $\bfs \in \calS$ with an undirected graph $(\calV,\calE)$ such that $\calV = \{s_1, s_2,\ldots, s_M\}$ and $\calE = \{(s_i, s_j) | \dist(s_i, s_j)\leq d, i\neq j \}$. Define the set of \emph{one-hop neighbors} of agent $i$ as $\calN_i:= \{j | (s_i, s_j)\in \calE\}\cup \{i\}$. We make the following assumption about the agents' motion.

\begin{assumption}
\label{ass:assumption3}
The distance between two consecutive states, $s_i(t)$ and $s_i(t+1)$, of agent $i$ is bounded, i.e., $\dist(s_i(t), s_i(t+1))\leq\epsilon$, for some $\epsilon > 0$.
\end{assumption}

This assumption is satisfied in many problems where, e.g., due to physical constraints, the agent states can only change by a bounded amount in a single time step. 

Define the \emph{potential neighbors} of agent $i$ at time $t$ as $\calP_i(t) := \{j|\dist(s_j(t), s_i(t))\leq2\epsilon+d\}$, which captures the set of agents that may become one-hop neighbors of agent $i$ at time $t+1$. Denote the joint state and action of the one-hop neighbors of agent $i$  by $\bfs_{\calN_{i}} = (s_{j_1},\ldots, s_{j_{|\calN_{i}|}})$ and $\bfa_{\calN_{i}} = (a_{j_1},\ldots, a_{j_{|\calN_{i}|}})$, respectively, where  $j_1,\ldots,j_{|\calN_{i}|} \in \calN_{i}$. Our key idea is to let agent $i$'s policy, $a_i = \mu_i(\bfs_{\calN_i})$, only depend on the one-hop neighbor states $\bfs_{\calN_i}$ instead of all agent states $\bfs$\revision{, and we assume that each agent can obtain its one-hop neighbor states through observation or communication. }
The intuition is that agents that are far away from agent $i$ at time $t$ have little impact on its current action $a_i(t)$. To emphasize that the output of a function $f: \prod_{i \in [M]}\calS_i \mapsto \bbR$ is affected only by a subset $\calN \subseteq [M]$ of the input dimensions, we use the notation $f(\bfs)=f(\bfs_{\calN})$ for $\bfs \in \calS$ and $\bfs_{\calN} \in \prod_{i \in \calN}\calS_i$ in the remainder of the paper. We make two additional assumptions on the problem structure to ensure the validity of our policy model.

\begin{assumption} \label{ass:reward}
The reward of agent $i$ can be fully specified using its one-hop neighbor states $\bfs_{\calN_i}$ and actions $\bfa_{\calN_i}$, i.e., $r_i(\bfs,\bfa) = r_i(\bfs_{\calN_i},\bfa_{\calN_i})$.
\end{assumption}

\revision{This assumption can always be satisfied by setting $d$ to the full environment range. In this case, the one-hop neighbor reward assumption becomes the standard reward definition, which depends on states and actions of all agents and is applicable to general MARL problems. For environments with local reward models, a smaller distance value $d$ can be chosen based on the specific environment configuration. For example, in collision avoidance problems, an agent's reward may depend only on the states and actions of nearby agents that maintain a safe distance. In multi-agent networks or sensing problems, the one-hop neighbors can be those  within communication or observation range. Next, we make a similar assumption for agent $i$'s transition model.} 

\revision{
\begin{assumption}\label{ass:transition}
The transition model of agent $i$ depends only on $\bfa_{\calN_i}$ and states $\bfs_{\calN_{i}}$, i.e., $p_i\left(s_{i}(t+1) \mid \bfs_{\calN_{i}}(t), \bfa_{\calN_i}(t)\right).$
\end{assumption} }


The objective of each agent $i$ is to obtain an optimal policy $\mu_i^*$ by solving the following problem:
\begin{equation}
    \mu_i^*(\bfs_{{\calN}_i}) =  \arg\max_{a_i} \max_{\bfa_{-i}} Q_i^{*}(\bfs,\bfa),\label{eq:policy_equation}
\end{equation}
where $Q_i^{*}(\bfs,\bfa) := \max_{\bfmu} Q_i^{\bfmu}(\bfs,\bfa)$ is the optimal action-value (Q) function introduced in the previous section.

The goal of this paper is to develop a MARL algorithm that (i) utilizes policy and value representations that scale favorably with the number of agents $M$ and (ii) allows efficient training on a distributed computing system containing compute stragglers.

\section{Distributed multi-Agent Reinforcement Learning with One-hop Neighbors (DARL1N)}
\label{sec:method}
This section develops the DARL1N algorithm to solve the MARL problem with proximity-graph structure introduced in Sec.~\ref{sec:problem}. DARL1N considers the effect of the one-hop neighbors of an agent in representing its Q and policy functions, which allows updating the Q and policy function parameters using only local one-hop neighborhood transitions.

Specifically, the Q function of each agent $i$ can be expressed as a function of its one-hop neighbor states $\bfs_{\calN_i}$ and actions $\bfa_{\calN_i}$ as well as the states $\bfs_{\calN_i^-}$ and actions $\bfa_{\calN_i^-}$ of the remaining agents that are not immediate neighbors of $i$:
\begin{equation}
    Q_i^{\bfmu}(\bfs, \bfa) = Q_i^{\bfmu}(\bfs_{\calN_i}, \bfs_{\calN_i^-}, \bfa_{\calN_i}, \bfa_{\calN_i^-}).
\end{equation}
Inspired by the SAC algorithm \cite{qu2020scalable}, we approximate the Q value with a function $\tilde{Q}_i^{\bfmu}$ that depends only on one-hop neighbor states and actions:
\begin{align}
    &\tilde{Q}_i^{\bfmu}(\bfs_{\calN_i}, \bfa_{\calN_i}) \notag\\
    & = \!\!\sum_{\bfs_{\calN_i^-}, \bfa_{\calN_i^-}}\!\!w_i(\bfs_{\calN_i}, \bfs_{\calN_i^-}, \bfa_{\calN_i}, \bfa_{\calN_i^-})Q_i^{\bfmu}(\bfs_{\calN_i}, \bfs_{\calN_i^-}, \bfa_{\calN_i}, \bfa_{\calN_i^-})\notag
\end{align}
where the weights $w_i(\bfs_{\calN_i}, \bfs_{\calN_i^-},\bfa_{\calN_i}, \bfa_{\calN_i^-})>0$ satisfy $\sum_{\bfs_{\calN_i^-}, \bfa_{\calN_i^-}}w_i(\bfs_{\calN_i}, \bfs_{\calN_i^-}, \bfa_{\calN_i}, \bfa_{\calN_i^-}) = 1$. The approximation error is given in the following lemma.




\begin{lemma} \label{lemma:approximation_error}
\revision{If the absolute value of agent $i$'s reward is upper bounded as $|r_i(\bfs_{\calN_i},\bfa_{\calN_i})|\leq\bar{r}$, for some $\bar{r}>0$, the approximation error between $\tilde{Q}_i^{\bfmu}(\bfs_{\calN_i}, \bfa_{\calN_i})$ and $Q_i^{\bfmu}(\bfs, \bfa)$ is bounded as:
\[
|\tilde{Q}_i^{\bfmu}(\bfs_{\calN_i}, \bfa_{\calN_i})-Q_i^{\bfmu}(\bfs, \bfa)|\leq\frac{2\bar{r}\gamma}{1-\gamma}.
\]}
\end{lemma}

\begin{proof}
See Appendix \ref{sec:proof_lemma1}.
\end{proof}


We parameterize the approximated Q function $\tilde{Q}_i^{\bfmu}(\bfs_{\calN_i}, \bfa_{\calN_i})$ and the policy $\mu_i(\bfs_{\calN_i})$ by $\theta_{i}$ and $\phi_{i}$, respectively. To handle the varying sizes of $\bfs_{\calN_i}$ and $\bfa_{\calN_i}$, in the implementation, we set the input dimension of $\tilde{Q}_i^{\bfmu}$ to the largest possible dimension of $(\bfs_{\calN_i}, \bfa_{\calN_i})$, and apply zero-padding for agents that are not within the one-hop neighborhood of agent $i$. The same procedure is applied to represent $\mu_i(\bfs_{\calN_i})$. 

To learn the approximated Q function $\tilde{Q}_i^{\bfmu}$, instead of incremental on-policy updates to the Q function as in SAC \cite{qu2020scalable}, we apply off-policy temporal-difference learning with a buffer similar to MADDPG \cite{lowe2017multi}. The parameters $\theta_i$ of the approximated Q function  are updated by minimizing the following temporal difference error:
\begin{align}
\label{eq:q_neighbor_maddpg}
\mathcal{L}\left(\theta_{i}\right) & = \mathbb{E}_{(\bfs_{\calN_i}, \bfa_{\calN_i}, r_i, \{\bfs_{\calN_l^\prime}\}_{\forall l\in\calN_i^\prime})\sim \calD_i} \!\!\left[\left(\tilde{Q}^{\bfmu}_{i}\left(\bfs_{\calN_i}, \bfa_{\calN_i}\right)-y\right)^{2}\right]\nonumber\\ y&=r_{i}+\gamma \hat{Q}_{i}^{\hat{\bfmu}}\left(\bfs_{\calN_i^\prime}, \bfa_{\calN_i^\prime}\right)
\end{align}
where $\calD_i$ is a replay buffer for agent $i$ that contains information only from $\calN_i$ and $\calN^\prime_i$, the one-hop neighbors of agent $i$ at the current and next time steps, and the one-hop neighbors $\calN^\prime_l$ for $l \in \calN^\prime_i$. \revision{Data from the one-hop neighbors of the next-step one-hop neighbors $\calN^\prime_i$ are needed to compute the next-step one-hop neighbors actions $\bfa_{\calN_i^\prime}$.} To stabilize the training, a target Q function $\hat{Q}_{i}^{\hat{\bfmu}}$ with parameters $\hat{\theta}_{i}$ and a target policy function $\hat{\mu}_i$ with parameters $\hat{\phi}_{i}$ are used. The parameters $\hat{\theta}_{i}$ and $\hat{\phi}_{i}$ 
are updated using Polyak averaging: $\hat{\theta}_{i} = \tau \hat{\theta}_{i} + (1-\tau)\theta_{i},
\hat{\phi}_{i} = \tau \hat{\phi}_{i} + (1-\tau)\phi_{i}$
where $\tau$ is a hyperparameter. In contrast to MADDPG \cite{lowe2017multi}, the replay buffer $\mathcal{D}_i$ for agent $i$ only needs to store its local interactions $(\bfs_{\calN_i}, \bfa_{\calN_i}, r_i, \{\bfs_{\calN_l^\prime}\}_{\forall l\in\calN^\prime_i})$ with nearby agents, \revision{where $\{\bfs_{\calN_l^\prime}\}_{\forall l\in\calN^\prime_i}$ is required to calculate $\bfa_{\calN_i^\prime}$.} Also, in contrast to SAC \cite{qu2020scalable}, each agent $i$ only needs to collect its own training data by simulating local two-hop interactions, \revision{which reduces agents' experience correlations} and allows efficient distributed training as explained in Sec. \ref{sec:training_architecture}.

Agent $i$'s policy parameters $\phi_{i}$ are updated using a gradient
\begin{equation}\label{eq:p_neighbor_maddpg}
\scaleMathLine{ \bfg(\phi_i)=\mathbb{E}_{\substack{\mathbf{s}_{\calN_i}, \bfa_{\calN_i} \sim \calD_i}}[\nabla_{\phi_{i}} \mu_{i}\left(\bfs_{\calN_{i}}\right) \nabla_{a_{i}} \tilde{Q}^{\bfmu}_{i}\left(\mathbf{s}_{\calN_i}, \bfa_{\calN_i}\right)],}
\end{equation}
where again data $\mathcal{D}_i$ only from local interactions is needed.

To implement the parameter updates proposed above, agent $i$ needs training data $\calD_i = (\bfs_{\calN_i}, \bfa_{\calN_i}, r_i, \{\bfs_{\mathcal{N}^\prime_l}\}_{l \in \calN'_i})$ from its one-hop neighbors at the current and next time steps. The relation between one-hop neighbors at the current and next time steps is captured by the following proposition.

\begin{proposition}
\label{pro:proposition1}
Under Assumption~\ref{ass:assumption3}, if an agent $j$ is not a potential neighbor of agent $i$ at time $t$, i.e., $j\not \in\mathcal{P}_i(t)$, it will not be a one-hop neighbor of agent $i$ at time $t+1$, i.e., $j\not \in\mathcal{N}_i(t+1)$.
\end{proposition}

\begin{proof}
See Appendix~\ref{sec:proof_proposition_1}.
\end{proof}

Proposition~\ref{pro:proposition1} allows us to decouple the global interactions among agents and limit the necessary observations to be among one-hop neighbors. To collect training data, at each time step, agent $i$ first interacts with its one-hop neighbors to obtain their states $\bfs_{\calN_i}$ and actions $\bfa_{\calN_i}$, and compute its reward $r_i(\bfs_{{\calN}_i}, \bfa_{\calN_i})$. To obtain $\bfs_{\calN_l^\prime}$ for all $l\in\calN^\prime_i$, we first determine agent $i$'s one-hop neighbors at the next time step, $\calN^\prime_i$. Using Proposition~\ref{pro:proposition1}, we let each potential neighbor $k \in \calP_i$ perform a transition to a new state $s^\prime_{k} \sim p_k(\cdot|\bfs_{{\calN}_k},\bfa_{{\calN}_k})$, which is sufficient to determine $\calN^\prime_i$. Then, we let the potential neighbors $\calP_l$ of each new neighbor $l \in \calN^\prime_i$ perform transitions to determine $\calN_l^\prime$ and obtain $\bfs_{\calN_l^\prime}$.

\begin{figure}[t]
    \centering
    \begin{minipage}{.4\linewidth}
    \phantom{{\small (a)~}}\includegraphics[width=0.75\linewidth]{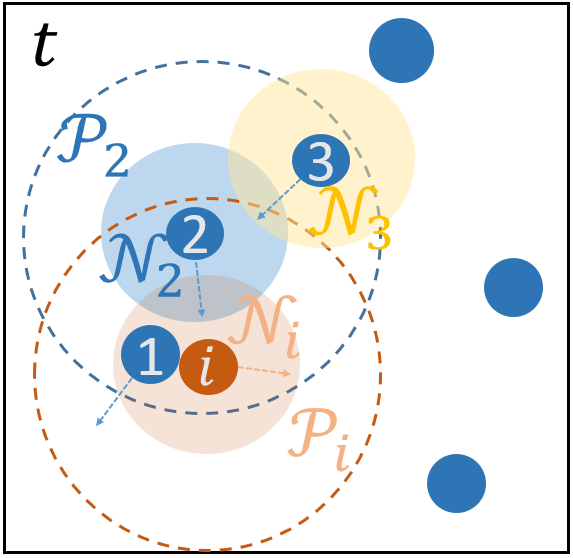}\\
    {\vspace{-0.4em}\small (a)~}\includegraphics[width=0.75\linewidth]{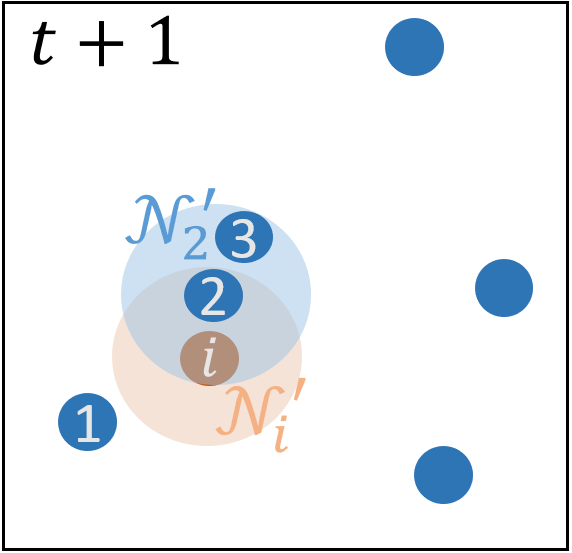}
    \end{minipage}%
    \hfill%
    \begin{minipage}{.6\linewidth}
    {\small (b)}\includegraphics[width=0.95\linewidth]{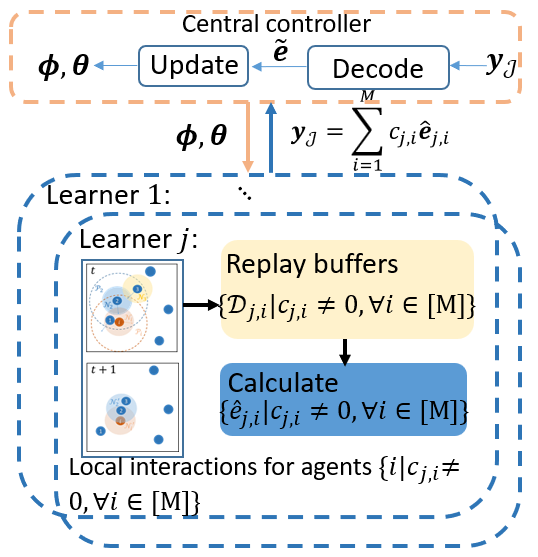}
    \vspace{-0.4em}
    \hspace{-0.8em}
    \end{minipage}
    \caption{(a) One-hop neighbor transitions from one time step to the next in a $d$-disk proximity graph; (b) Coded distributed learning architecture.}
    \label{fig:distributed_learning}
    \vspace{-1.5em}
\end{figure}

Fig.~\ref{fig:distributed_learning}(a) illustrates the data collection process. At time $t$, agent $i$ obtains $\bfs_{\calN_i}$, $\bfa_{\calN_i}$, and $r_i(\bfs_{\calN_i}, \bfa_{\calN_i})$ for $\calN_i = \{i, 1\}$. Then, the potential neighbors of agent $i$, $\calP_i=\{1,2,i\}$, proceed to their next states at time $t+1$. This is sufficient to determine that $\calN_i^\prime= \{i, 2\}$ and obtain $\bfs_{\calN^\prime_i}$. Finally, we let agent $3$, which belongs to set $\calP_2=\{i, 1, 2, 3\}$, perform a transition to determine that $\calN_2^\prime= \{i, 2, 3\}$ and obtain $\bfs_{\calN^\prime_2}$.

As each agent only needs to interact with one-hop neighbors to update its parameters, the agents can be trained in parallel on a distributed computing architecture, where each compute node only needs to simulate the two-hop neighbor transitions for agents assigned to it for training.

\section{Coded Distributed Learning Architecture}
\label{sec:training_architecture}
In this section, we introduce an efficient and resilient distributed learning architecture for training DARL1N. A coded distributed learning architecture, illustrated in Fig.~\ref{fig:distributed_learning}(b), consists of a central controller and $N$ compute nodes, called \emph{learners}. The central controller stores a copy of all parameters of the policy $\phi_{i}$, target policy $\hat{\phi}_{i}$, Q function $\theta_{i}$, and target Q function $\hat{\theta}_{i}$, for all $i \in [M]$. In each training iteration, the central controller broadcasts all agents' parameters to all learners, who then calculate and return the gradients required for updating the parameters. 
In a traditional uncoded distributed learning architecture, each agent is only trained (with its policy and value gradients computed) by a single learner.
If any learner becomes slow or unresponsive, i.e., a straggler, the whole training procedure is delayed or may fail. Our coded distributed learning architecture addresses the possible presence of stragglers in the computing system by introducing redundant computations. We let more than one learner train each agent, which not only improves the system resilience to stragglers but also accelerates the training speed, as we show in Sec. \ref{sec:coded_darl1n_exp}. To describe which learners are assigned to train each agent, we introduce an assignment matrix $\bfC\in\mathbb{R}^{N\times M}$ with non-zero 
entries $c_{j,i}\neq 0$ indicating that learner $j\in [N]$ is assigned to train agent $i\in [M]$. The complete set of learners assigned to train an agent $i$ can then be determined by $\{j|c_{j,i}\neq 0,\forall j \in [N]\}$. To construct the assignment matrix $\bfC$, we apply coding theory as explained in Sec.\ref{sec:assignment}.

To calculate the gradients for an agent $i$, each learner $j$ with $c_{j,i}\neq 0$ simulates transitions to get the interaction data $(\bfs_{\calN_i}, \bfa_{\calN_i}, r_i, \{\bfs_{\calN_l^\prime}\}_{l\in\calN^\prime_i})$  as described in Sec. \ref{sec:method}, which are stored in a replay buffer $\calD_{j, i}$. After that, learner $j$ calculates the gradients of the temporal difference error needed for updating the Q function parameters $\theta_i$ of agent $i$ using \eqref{eq:q_neighbor_maddpg} and updating the policy parameters $\phi_i$ using \eqref{eq:p_neighbor_maddpg}.

As the replay buffer $\calD_{j,i}$ can have a large size, 
to improve efficiency, we use a mini-batch $\calB_{j,i}$ uniformly sampled from $\calD_{j,i}$ to estimate the expectations in \eqref{eq:q_neighbor_maddpg}-\eqref{eq:p_neighbor_maddpg}. In particular, the temporal difference error in \eqref{eq:q_neighbor_maddpg} is estimated with:
\begin{align}
\label{eq:Q_gradient_estimated}
\hat{\mathcal{L}_j}\left(\theta_{i}\right) & = \frac{1}{|\calB_{j,i}|}\hspace{-3.7em}\sum_{\substack{(\bfs_{\calN_i}, \bfa_{\calN_i}, r_i, \{\bfs_{\calN_l^\prime}\}_{\forall l\in\calN_i^\prime})\\\in \calB_{j,i}}}\hspace{-2.5em} \!\!\left(\tilde{Q}^{\bfmu}_{i}\left(\bfs_{\calN_i}, \bfa_{\calN_i}\right)-y\right)^{2}\nonumber\\ y&=r_{i}+\gamma \hat{Q}_{i}^{\hat{\bfmu}}\left(\bfs_{\calN_i^\prime}, \bfa_{\calN_i^\prime}\right).
\end{align}
Similarly, the gradients used to update policy parameters are estimated with:
\begin{align}
\label{eq:policy_gradient_estimated}
\hat{\bfg}_j(\phi_i)=\frac{1}{|\calB_{j,i}|}\hspace{-2.1em}\sum\limits_{(\bfs_{\calN_i},\bfa_{\calN_i})\in\calB_{j,i}}\hspace{-1.5em}\nabla_{\phi_{i}} \mu_{i}\left(\bfs_{\calN_{i}}\right) \nabla_{a_{i}} \tilde{Q}^{\bfmu}_{i}\left(\mathbf{s}_{\calN_i}, \bfa_{\calN_i}\right).
\end{align}
Let $\hat{\bfe}_{j,i}=[\nabla \hat{\calL}_j(\theta_i), \hat{\bfg}_j(\phi_i)]$ 
denote the concatenation of estimated gradients. Instead of directly returning the estimated gradients for all agents trained by learner $j$, i.e., $\{\hat{\bfe}_{j,i}| \forall i\in [M], c_{j,i}\neq 0\}$, learner $j$ calculates a linear combination of the gradients: $y_j =\sum_{i=1}^{M}c_{j,i}\hat{\bfe}_{j,i}$
with weights provided by the assignment matrix $\bfC$ and returns $y_j$ back to the central controller.

At the central controller, let $\bfy_\calJ$ denote the results that have arrived by a certain time from learners $\calJ=\{j|y_j$ is received$\}$. Moreover, let $\bfC_{\calJ}\in\mathbb{R}^{|\calJ|\times M}$ be a submatrix of $\bfC$ formed by the $j$-th row of $\bfC,\forall j\in \calJ$. The received gradients $\bfy_\calJ$ satisfy:
\begin{align}
    \bfy_\calJ &= \bfD\bfq \nonumber\\
    \bfq &= [\hat{\bfe}_{1,1}; \hat{\bfe}_{1,2}; \ldots;\hat{\bfe}_{N,M-1}; \hat{\bfe}_{N,M}]\label{eq:grad_estimator}
\end{align}
where $\bfD\in\mathbb{R}^{|\calJ|\times MN}$ is constructed as follows: for $i$-th row of $\bfD$, fill the $(1 + (\calJ_i-1)M)$-th to $(\calJ_iM)$-th entries with $i$-th row of $\bfC_{\calJ}$ and set all other entries to 0, $\calJ_i$ denotes $i$-th element of $\calJ$. 
The vector $\bfq$ is a concatenation of all the gradients estimated by all learners. The central controller updates the agents' parameters once it receives enough results to decode all estimated gradients, denoted as $\tilde{\bfe}$. This happens when $\rank(\bfC_\calJ) = M$, and the decoding equation is given as follows:
\begin{equation}
\label{eq:recover}
    \tilde{\bfe} = (\bfC_\calJ^T\bfC_\calJ)^{-1}\bfC^T_\calJ\bfy_\calJ.
\end{equation}
Alg.~\ref{alg:darl1n} summarizes the coded training procedure of DARL1N over a distributed computing architecture, referred to as the Coded DARL1N.

In Coded DARL1N, the gradients $\tilde{\bfe}$ used by the central controller for parameter updates are stochastic gradients computed by mini-batch samples, which are estimates of the true gradients $\bfe=[\bfe_1,\ldots,\bfe_M]$, where $\bfe_i = [\nabla \calL(\theta_i), \bfg(\phi_i)]$  with $\calL(\theta_i)$ and $\bfg(\phi_i)$ defined in \eqref{eq:p_neighbor_maddpg} and \eqref{eq:q_neighbor_maddpg}, respectively. The estimation performance is illustrated in the following theorem.

\begin{theorem}
\label{theorem:convergence}
The mini-batch stochastic gradients $\tilde{\bfe}$ computed by Coded DARL1N are unbiased estimates of the true gradients $\bfe$, with variance determined by the assignment matrix $\bfC$.
\end{theorem}
\begin{proof}
See Appendix \ref{sec:proof_proposition_convergence}.
\end{proof}

Based on Theorem \ref{theorem:convergence}, we can infer that Coded DARL1N converges asymptotically similarly to 
other stochastic gradient descent-based methods \cite{qian2020impact}.

\begin{algorithm}[t]
    \small
    \tcp{\textbf{Central controller:}}
     Initialize policy, target policy, Q, and target Q parameters $\bfphi=\{\phi_{i},\hat{\phi}_{i}\}_{i\in[M]}$, $\bftheta = \{\theta_{i}, \hat{\theta}_{i}\}_{i\in[M]}$;\\
        Broadcast $\bfphi, \bftheta$ to the learners;\\
        $\bfy_{\calJ} \leftarrow [ \ ]$; \\
        \Do{ $\tilde{\bfe}$ is not recoverable}
        {
        Listen to channel and
        collect $y_j$ from the learners: $\bfy_{\calJ}\leftarrow[\bfy_{\calJ}, y_j],j\in[N]$;
        }
        Send acknowledgements to learners; \\
        Update $\bfphi, \bftheta$ with $\tilde{\bfe}$;\\
    \tcp{\textbf{Learner $j$:}}
    \mbox{Initialize replay buffer $\calD_{j,i}$;}\\
    \For{$iter = 1:\text{max\_iteration}$}
    {
    Listen to channel; \\
    \If{$\bfphi, \bftheta$ received from the central controller}
    { 
    $y_j\leftarrow0$; $i \leftarrow 1$;\\
    \While{$i \leq M$ and no acknowledgement received}{
    \If{$c_{j,i}\not = 0$}
    {
        \tcp{\textbf{Local interactions:}}
        Perform local interactions to collect training data for agent $i$ and store the data into $\calD_{j,i}$;\\
        Sample a mini-batch from $\mathcal{D}_{j, i}$ and calculate $\hat{\bfe}_{j, i}$ using \eqref{eq:Q_gradient_estimated}-\eqref{eq:policy_gradient_estimated};}
        $y_j\leftarrow y_j + c_{j,i}\hat{\bfe}_{j,i}$;\\
        $i \leftarrow i + 1$;\\
        Send updated $y_j$ to the central controller;
        }}}
\caption{Coded DARL1N} \label{alg:darl1n}
\end{algorithm}

\section{Assignment Matrix Construction and Assessment}
\label{sec:assignment}
The assignment matrix $\bfC$ affects both the policy variance and computational efficiency of Coded DARL1N.  In this section, we explore different schemes, both uncoded and coded, for constructing the assignment matrix, and \revision{conduct theoretical analyses on their performance.}
\subsection{Assignment Matrix Construction}

\subsubsection{Uncoded Assignment Scheme}
In an uncoded distributed training architecture, different learner nodes train different agents exclusively. The assignment matrix can then be constructed as:
$\bfC^{\text{Uncoded}}=[\bfI_M|\mathbf{0}]^T$, where $\bfI_{M}\in \mathbb{R}^{M\times M}$ is an identity matrix.

\subsubsection{Coded Assignment Schemes}
Coded distributed training assigns each agent to multiple learners. Here, we investigate five codes, where the encoding matrices can be directly utilized as the assignment matrix. 
\revision{
\begin{itemize}
    \item \textit{MDS Code}: An MDS code \cite{lacan2004systematic} is an erasure code with the property that any square submatrix of its encoding matrix $\bfC^{\text{MDS}}$ has full rank. A Vandermonde matrix \cite{klinger1967vandermonde,wang2021coding} is commonly used for encoding,
    with the $(j,i)$-th entry given by $\bfC_{j,i}=\alpha_{i}^{j-1}$, where $\alpha_i\neq 0$, $i \in [M]$, can be any non-zero distinct real numbers.
    
    \item \textit{Random Sparse Code}: Compared to an MDS code, a Random Sparse code \cite{lee2017coded} results in a sparser assignment matrix with the $(j,i)$-th entry given by:
        \begin{equation}
        \bfC^{\text{Random}}_{j,i}  = \begin{cases} 0 , & \text{with probability}  \ 1 - \xi ,\\ \zeta, & \text{with probability} \ \xi. \end{cases} \label{eq:C_Sparse}
        \end{equation}
    where $\zeta\sim\calN(0, 1)$, $\xi\in [0,1]$. 
    
    \item \textit{Repetition Code}: A Repetition code  \cite{lee2017coded} assigns agents to the learners repetitively in a round-robin fashion. The $(j,i)$-th entry of the assignment matrix is given by:
    \begin{equation}
    \bfC^{\text{Repetition}}_{j,i}  = \begin{cases} 1, & \text{if} \ i = (j\text{~mod~}M),\\ 0, & \text{else}, \end{cases} \label{eq:C_repetition}
    \end{equation} 
    where $\text{mod}$ is the modulo operator.
    
    \item \textit{LDGM Code}: An LDGM code \cite{cai2021systematic} is a special type of a Low Density Parity Check code \cite{gallager1962low} that constructs a sparser assignment matrix. By applying a systematic biased random code ensemble \cite{cai2021systematic}, the LDGM assignment matrix takes the form: $\bfC^{\text{LDGM}} = [\bfI_{M}|\hat{\bfP}]^T$
    , where each entry of $\hat{\bfP}$ is generated independently according to a Bernoulli distribution with success probability $\text{Pr}(\hat{\bfP}_{i,j}=1)=\rho$. Note that when $\rho\leq\frac{1}{2}$, the assignment matrix of LDGM code has a low density.
\end{itemize}
}

\subsection{Analysis and Comparison of Assignment Schemes}

\revision{We provide a theoretical analysis and comparison of different assignment schemes from the following aspects: 1) \textit{computation overhead} and 2) \textit{resilience to stragglers}.}

\subsubsection{Computation Overhead}
\revision{
The coded schemes mitigate the impact of stragglers by assigning each agent to multiple learners. The training performed by the extra learners is redundant. To quantify the \emph{computation overhead} introduced by this redundancy, we use the following metric:
\begin{equation}
\label{eq:computation_overhead}
    o_{c} = \frac{1}{M} \sum_{j=1}^{N}\sum_{i=1}^{M}\mathds{1}_{\bfC_{j,i}\neq0} - 1, \nonumber
\end{equation}
where the first term on the right hand side calculates the average number of learners used for training each agent, and $o_c \geq 0$. 
Using the above metric, the computation overhead of each assignment scheme can be derived as follows:
\begin{itemize}
    \item \textit{Uncoded}: $o_c^{\text{Uncoded}} = 0$, 
    as each agent is assigned to only one learner in the uncoded scheme.
    \item \textit{MDS Code}: $o_c^{\text{MDS}} = N-1$.  
    All entries of the MDS assignment matrix are non-zero, indicating that each agent is assigned to all learners.
    \item \textit{Random Sparse Code}: 
    $o_c^{\text{Random}}$ depends on the parameter $\xi$, but its expectation is derived as $\mathbb{E}(o_c^{\text{Random}}) = \xi N-1$.
    \item \textit{Repetition Code}: $o_c^{\text{Repetition}} = \frac{N}{M}-1$.
    \item \textit{LDGM Code}: 
    $\mathbb{E}(o_c^{\text{LDGM}}) = (N-M)\rho$.
\end{itemize}
Among these schemes, the MDS code incurs the highest computation overhead, while the uncoded scheme results in the lowest. The overhead introduced by 
the Random Sparse and LDGM codes depend on their parameters, $\xi$ and $\rho$.
}

\subsubsection{Resilience to Stragglers}

\revision{
According to \eqref{eq:recover}, the central controller can update the agents' gradients only after receiving results from enough learners, specifically when $\rank(\bfC_{\calJ}) = M$. 
To evaluate the resilience of assignment schemes to stragglers, we analyze the probability of each scheme being influenced by stragglers under the following assumption. 
\begin{assumption}
\label{assumption:straggler}
In each training iteration, each compute node in a distributed computing system has a probability of $\eta \in [0,1]$ to become a straggler.
\end{assumption}}

\revision{
The Random Sparse and LDGM codes have randomly generated entries that depend on parameters $\xi$ and $\rho$, making them hard to analyze theoretically. We focus our analysis on the Uncoded, MDS, and Repetition codes as follows.

%
    \begin{proposition}
    \label{prop:uncoded}
     The probability that the Uncoded scheme 
will be influenced by stragglers is $1-(1-\eta)^M$.
    \end{proposition}
    \begin{proposition}
    \label{prop:MDS}
     The probability that the MDS code will be influenced by stragglers is $\sum_{j=N-M+1}^{N} {N\choose j}(1-\eta)^{N-j}\eta^j$.
    \end{proposition}
        \begin{proposition}
    \label{prop:rep} 
     The probability that the Repetition code will be influenced by stragglers is $1-(1-\eta^{\frac{N}{M}})^M$ given that $\frac{N}{M}$ is positive integer.
    \end{proposition}}

    \revision{
    The proof of Proposition \ref{prop:uncoded} is direct since the Uncoded scheme is affected by any straggler. The proofs of Propositions \ref{prop:MDS} and \ref{prop:rep} are provided in Appendices~\ref{sec:proof_proposition_3} and \ref{sec:proof_proposition_5}, respectively.
    %
}

\section{Experiments}
\label{sec:experiments}
In this section, we evaluate the DARL1N algorithm and our coding schemes for mitigating \revision{compute} stragglers.

\subsection{Performance of DARL1N}
We conduct a series of comparisons between DARL1N and four state-of-the-art MARL algorithms. For fair comparison, we train DARL1N using a distributed learning architecture with uncoded assignments, and \revision{run our experiments on the Amazon EC2 computing clusters \cite{amazonEC2}, which are considered reliable and free of stragglers.}

\subsubsection{Experiment Settings}

\paragraph{Environment Configurations} We evaluate DARL1N in five environments, including the Ising Model \cite{yang2018mean}, Food Collection, Grassland, Adversarial Battle \cite{long2020evolutionary}, and Multi-Access Wireless Communication \cite{qu2020scalable}, which cover cooperative and mixed cooperative competitive games. Please refer to \cite{qu2020scalable,yang2018mean, long2020evolutionary} for the description of each environment.

To understand the scalability of our method, we
vary the number of agents $M$ and the size of the local state spaces. The specific configurations for the first four environments can be referenced in the conference version \cite{wang2022darl1n}. 
In the Multi-Access Wireless Communication environment, which was not considered in \cite{wang2022darl1n}, we adopt the setting in \cite{qu2020scalable} and consider a grid of $3\times 3$ agents, with each having a state space of $\calS_i=\{0,1\}^z$ 
to indicate whether there is a packet to send by time step $z$, where $z$ is set to either $z=2$ or $z=10$.


\paragraph{Neighborhood Configuration}

\revision{In both the Ising Model and Multi-Access Wireless Communication environments, the agents are arranged in a two dimensional lattice graph, with rewards depending solely on their proximal agents. Consequently, an agent's one-hop neighbors are naturally defined as those directly connected to it, including itself. 
In the other three environments, 
the agents are trained to avoid one another. Therefore, the one-hop neighbor distances $d$ are naturally set as the Euclidean safety distances. Specifically, 
the safety distances (or $d$) are set to $0.15, 0.2, 0.25, 0.3, 0.35$ when $M=3, 6, 12, 24, 48$, respectively. 
Each agent observes its one-hop neighbors to obtain one-hop neighbor states. Other distance metrics that account for velocity can be employed, which is left for future work. 
}

\paragraph{Benchmarks} 
We compare our method with four state-of-the-art MARL algorithms: MADDPG \cite{lowe2017multi}, MFAC \cite{yang2018mean}, EPC \cite{long2020evolutionary}, and SAC \cite{qu2020scalable}. The SAC algorithm only works in the Multi-Access Wireless Communication environment due to the reward assumption. 

\paragraph{Evaluation Metrics} We measure the performance using two criteria: \textit{training efficiency} and \textit{policy quality}. To measure the training efficiency, we use two metrics: 1) \textit{average training time} spent to run a specified number of training iterations and 2) \textit{convergence time}.  The convergence time is defined as the time when the variance of the average total training reward over  90 consecutive iterations does not exceed 2\% of the absolute mean reward, where the average total training reward is the total reward of all agents averaged over 10 episodes in three training runs with different random seeds. To measure policy quality, we use \textit{convergence reward}, which is the average total training reward at the convergence time.

\begin{figure*}[t]
    \centering
    {\small (a)}\subfigure{\label{fig:ising_tt}\includegraphics[width=0.25\linewidth]{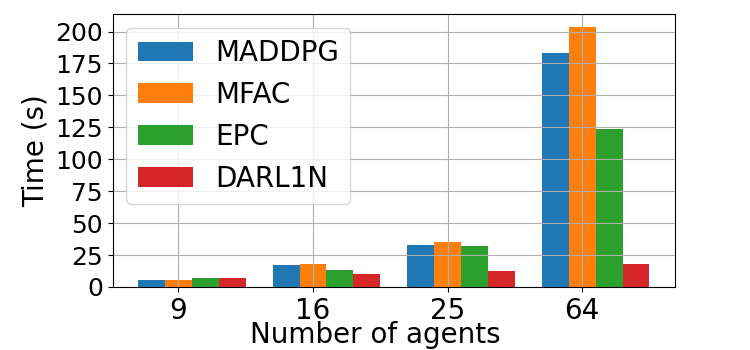}\hspace*{-1.8em}}
    {\small (b)}\subfigure{\label{fig:spread_tt}\includegraphics[width=0.25\linewidth]{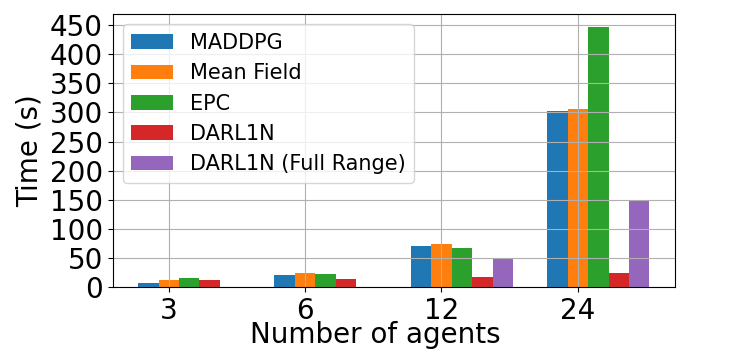}\hspace*{-1.8em}}
    {\small (c)}\subfigure{\label{fig:grass_tt}\includegraphics[width=0.25\linewidth]{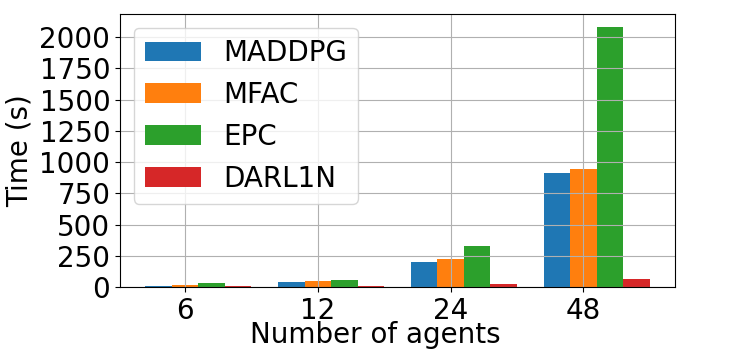}\hspace*{-1.8em}}
    {\small (d)}\subfigure{\label{fig:adv_tt}\includegraphics[width=0.25\linewidth]{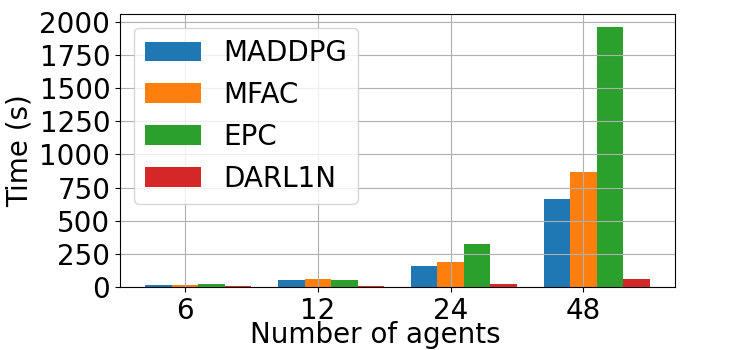}}
    \hspace*{-2.0em}
    \vspace{-0.2cm}
	\caption{Average training time of different methods to run (a) 10 iterations in the Ising Model, (b) \revision{30 iterations in the Food Collection}, (c) 30 iterations in the Grassland, and (d) 30 iterations in the Adversarial Battle environments.}
	\label{fig:training_time}
 \vspace{-0.7cm}
\end{figure*}

\paragraph{Computing Configurations}
The computing resources are configured in a way so that DARL1N utilizes roughly the same or fewer resources than the baseline methods, as described in \cite{wang2022darl1n}. For the Multi-Access Wireless Communication environment,  we employ the Amazon EC2 $c5n.large$ instance for DARL1N training, the $z1d.3xlarge$ instance for MADDPG and MFAC, and the $c5.12xlarge$ instance for EPC training.  The configurations for the training parameters, as well as the representations of policy and Q functions can be found in \cite{wang2022darl1n}.

\subsubsection{Comparative Studies}
\paragraph{Ising Model}
\setlength{\tabcolsep}{0.3pt}
\begin{table}
\caption{Convergence time and convergence reward of different methods in the Ising Model environment.}
\label{tab:ising_conv}
\scriptsize\centering{\begin{tabular}{|c|c|c|c|c|c|c|c|c|}
	\hline
	\bf \multirow{2}{*}{Method} & \multicolumn{4}{c}{Convergence Time (s)}& \multicolumn{4}{|c|}{Convergence Reward}\\
	\cline{2-9}
	& $M=9$ & $M=16$ & $M=25$ & $M=64$ & $M=9$ & $M=16$ & $M=25$ & $M=64$\\
	\hline
    MADDPG & 62 & 263 & 810 & 1996 & 460 & 819 & \textbf{1280} & 1831 \\
    \hline
    MFAC & 63 & 274 & 851 & 2003 & \textbf{468} & 814 & 1276 & 1751\\
    \hline
    EPC  & 101 & \textbf{26} & \textbf{51} & \textbf{62} & \textbf{468} & \textbf{831} &1278 & \textbf{3321}\\
    \hline
    EPC Scratch & 101 & 412 & 993 & 2995 & \textbf{468} & 826 & 1275 & 2503\\
    \hline
    \textbf{DARL1N}  & \textbf{38} & 102 & 210 & 110 & 465 & 828 & 1279 & 2282\\
    \hline
\end{tabular}}
\vspace{-2.0em}
\end{table}

As shown in Tab.~\ref{tab:ising_conv}, when the number of agents is small ($M=9$), 
all methods achieve roughly the same reward. DARL1N takes the least amount of time to converge while EPC takes the longest time. When the number of agents increases, it can be observed that the EPC converges immediately and the convergence reward it achieves when $M = 64$ is much higher than the other methods. The reason is that, in the Ising Model, each agent only needs information of its four fixed neighbors, and hence in EPC the policy obtained from the previous stage can be applied to the current stage. The other methods train the agents from scratch without curriculum learning. 
For illustration, we also show the 
performance achieved by training EPC from scratch without curriculum learning (labeled as EPC Scratch in Tab.~\ref{tab:ising_conv}). The results show that EPC Scratch converges much slower than EPC as the number of agents increases. Note that when the number of agents is 9, EPC and EPC Scratch are the same. Moreover, DARL1N achieves a reward comparable with that of EPC Scratch but converges much faster. 
From Fig.~\ref{fig:ising_tt}, we can observe that DARL1N requires much less time to perform a training iteration than the benchmark methods. 

\paragraph{Food Collection}

\setlength{\tabcolsep}{1.0pt}
\begin{table}
\caption{Convergence time and convergence reward of different methods in the Food Collection environment.}
\label{tab:food_conv}
\scriptsize\centering
\begin{tabular}{|c|c|c|c|c|c|c|c|c|}
	\hline
	\bf \multirow{2}{*}{Method} & \multicolumn{4}{c}{Convergence Time (s)}& \multicolumn{4}{|c|}{Convergence Reward}\\
	\cline{2-9}
	& $M=3$ & $M=6$ & $M=12$ & $M=24$ & $M=3$ & $M=6$ & $M=12$ & $M=24$\\
	\hline
     MADDPG & \textbf{501} & 1102 & 4883 & 2005 & 24 & 24 & -112 & -364  \\
     \hline
      MFAC & 512 & 832 & 4924 & 2013 & 20 & 23& -115 & -362 \\
      \hline
    EPC  & 1314 & 723 & 2900 & 8104 & \textbf{31} & \textbf{34} & -16 & -87\\
    \hline
    \textbf{DARL1N}  & 502 & \textbf{382} & \textbf{310} & \textbf{1830} & 14 & 25 & \textbf{13} & \textbf{-61}\\
\hline
\end{tabular}
\vspace{-2.0em}
\end{table}

\begin{figure}[t]
    \centering
    {\small (a)}\subfigure{\label{fig:spread_rewardc}\includegraphics[width=0.45\linewidth]{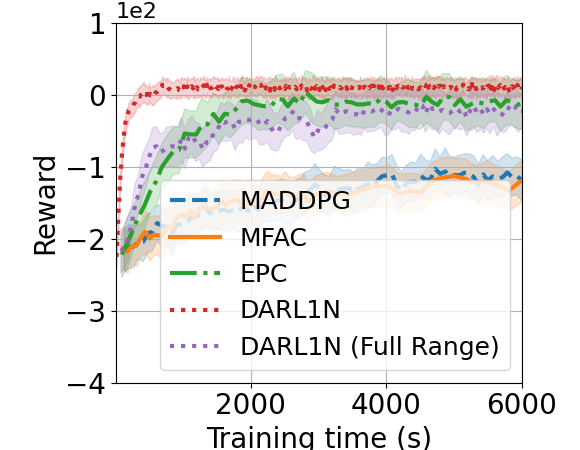}}%
    \hfill%
    {\small (b)}\subfigure{\label{fig:spread_rewardd}\includegraphics[width=0.45\linewidth]{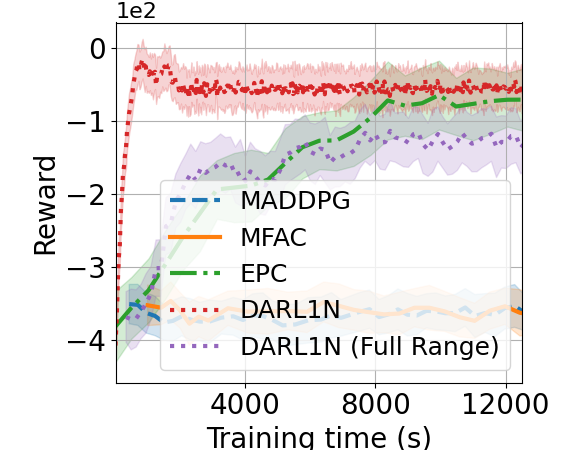}}
    \caption{\revision{Average total training reward of different methods in the Food Collection environment when there are (a) $M=12$, and (b) $M=24$ agents.}}
    \label{fig:spread_reward}
    \vspace{-0.5cm}
\end{figure}

Tab.~\ref{tab:food_conv} shows that, in Food Collection, when the problem scale is small, 
DARL1N, MADDPG and MFAC achieve similar performance in terms of policy quality. 
As the problem scale increases, the performance of MADDPG and MFAC degrades significantly and becomes much worse than DARL1N or EPC when $M=12$ and $M=24$, which is also illustrated in Fig.~\ref{fig:spread_reward}. The convergence reward achieved by DARL1N is comparable or sometimes higher than that achieved by EPC. Moreover, the convergence speed of DARL1N is the highest among all methods in all scenarios. \revision{Notably, the convergence time of DALR1N and EPC increases, while that of MADDPG and MFAC decreases as $M$ increases to 24. This occurs because MADDPG and MFAC fail to handle such large-scale networks, causing them to stop learning earlier.}

\revision{ 
To evaluate the impact of the proposed one-hop neighbor reward formulation on the learning performance, 
we also present in Fig.~\ref{fig:spread_reward} the training rewards of DARL1N with a standard reward definition, labeled as DARL1N (Full Range), whose neighbor distance $d$ is set to 
cover the entire environment, thereby including all agents as one-hop neighbors. The results show that DARL1N (Full Range) achieves performance comparable to EPC but performs worse than DARL1N with a small number of agents considered as one-hop neighbors. This suggests that in the Food Collection environment, agent behavior primarily depends on interactions with a nearby, smaller group of agents. 
Fig.~\ref{fig:spread_tt} illustrates the training time of DARL1N (Full Range), which increases compared to DARL1N due to the inclusion of more agents in the reward calculations.} 


Fig.~\ref{fig:spread_tt} also presents the training times of the four benchmarks. Among all methods compared, DARL1N achieves the highest training efficiency and its training time grows linearly as the number of agents increases. When $M=24$, EPC takes the longest time to train. This is because of the complex policy and Q neural network architectures in EPC, the input dimensions of which grow linearly and quadratically, respectively, with more agents.

\begin{figure}[t]
    \centering
    {\small (a)\hspace*{-0.05em}}\subfigure{\includegraphics[width=0.45\linewidth]{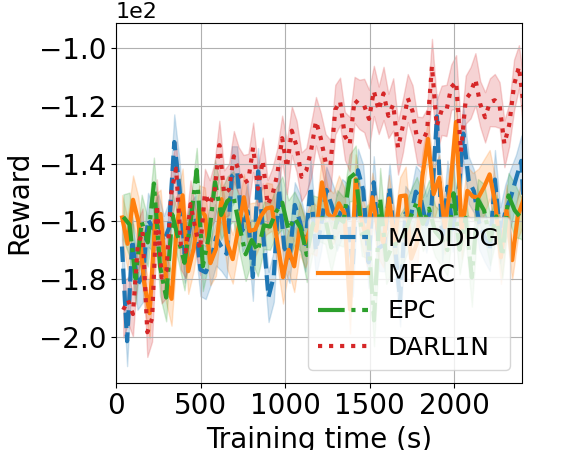}}\hspace*{-0.25em}
    {\small (b)\hspace*{-0.05em}}\subfigure{\includegraphics[width=0.45\linewidth]{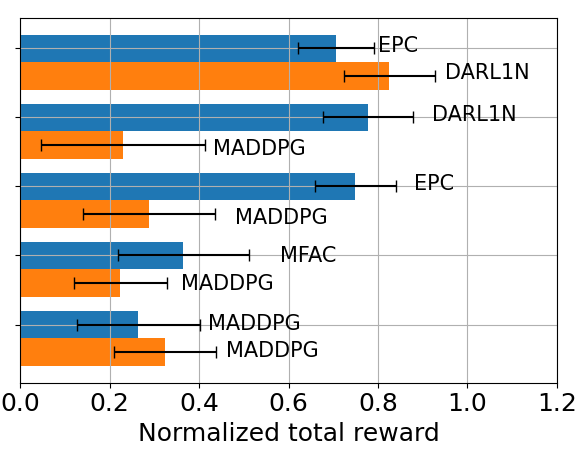}}\hspace*{-1em}
    \caption{\revision{(a) Average total training reward of different methods in the Food Collection environment with global reward and transition models when there are $M=8$ agents} and (b) Mean and standard deviation of normalized total reward of competing agents trained by different methods in the Adversarial Battle environment with $M=48$.}
    \label{fig:adv_score}
    \vspace{-1.95em}
\end{figure}

\revision{To demonstrate DARL1N's applicability to general MARL problems with global reward and transition models, we conduct a comparison study using a variant of the Food Collection environment where agents must coordinate to exclusively collect all the food. 
As shown in Fig.~\ref{fig:adv_score}(a), 
DARL1N achieves the highest reward level with the fastest training speed. 
This is due to its distributed learning architecture, which reduces training experience correlation and accelerates training through parallel computing, even without decomposition. In contrast, EPC performs significantly worse, likely because curriculum learning struggles with global agent coordination.}



\paragraph{Grassland}

\setlength{\tabcolsep}{0.5pt}
\begin{table}[t]
\caption{Convergence time and convergence reward of different methods in the Grassland environment.}
\label{tab:grassland_conv}
\scriptsize\centering
\begin{tabular}{|c|c|c|c|c|c|c|c|c|}
	\hline
	\bf \multirow{2}{*}{Method} & \multicolumn{4}{c}{Convergence Time (s)}& \multicolumn{4}{|c|}{Convergence Reward}\\
	\cline{2-9}
	& $M=6$ & $M=12$ & $M=24$ & $M=48$ & $M=6$ & $M=12$ & $M=24$ & $M=48$\\
	\hline
     MADDPG & 423 & 6271 & 2827 & 1121 & 21 & 11 & -302 & -612 \\
     \hline
      MFAC & 431 & 7124 & 3156 & \textbf{1025} & \textbf{23} & 9 & -311 & -608\\
      \hline
    EPC  & 4883 & 2006 & 3324 &15221 & 12 & 38 & 105 & 205\\
    \hline
    \textbf{DARL1N}  & \textbf{103} & \textbf{402} & \textbf{1752} & 5221 & 18 & \textbf{46} & \textbf{113} & \textbf{210}\\
\hline
\end{tabular}
\vspace{-1.2em}
\end{table}

Similar as the results in the Food Collection environment, the policy generated by DARL1N is equally good or even better than those generated by the benchmark methods, as shown in Tab.~\ref{tab:grassland_conv} and Fig.~\ref{fig:grass_tt}, especially when the problem scale is large. DARL1N also has the fastest convergence speed and takes the shortest time to run a training iteration.

\paragraph{Adversarial Battle}

\begin{table}[t]
\caption{Convergence time and convergence reward of different methods in the Adversarial Battle environment.}
\label{tab:adv_conv}
\scriptsize\centering
\begin{tabular}{|c|c|c|c|c|c|c|c|c|}
	\hline
	\bf \multirow{1}{*}{Method} & \multicolumn{4}{c}{Convergence Time (s)}& \multicolumn{4}{|c|}{Convergence Reward}\\
	\cline{2-9}
	& $M=6$ & $M=12$ & $M=24$ & $M=48$ & $M=6$ & $M=12$ & $M=24$ & $M=48$\\
	\hline
     MADDPG & 452 & 1331 & 1521 & 7600 & -72 & -211 & -725 & -1321 \\
     \hline
      MFAC & 463 & 1721 & 1624 & 6234 & -73 & -221 & -694 & -1201\\
      \hline
    EPC  & 1512 & 1432 & 2041 & 9210 & -75 & -215 & \textbf{-405} &\textbf{-642}\\
    \hline
    \textbf{DARL1N}  & \textbf{121} & \textbf{756} & \textbf{1123} & \textbf{3110} & \textbf{-71} & \textbf{-212} & -410 & -682\\
    \hline
\end{tabular}
\vspace{-2.5em}
\end{table}

\begin{figure}[t]
    \centering
    {\small (a)}\subfigure{\includegraphics[width=0.45\linewidth]{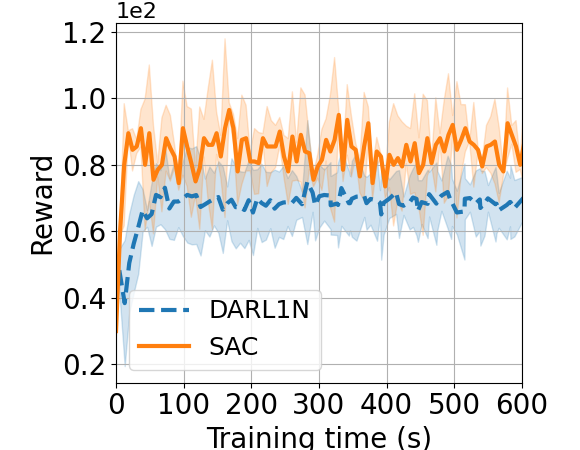}}%
    \hfill%
    {\small (b)}\subfigure{\includegraphics[width=0.45\linewidth]{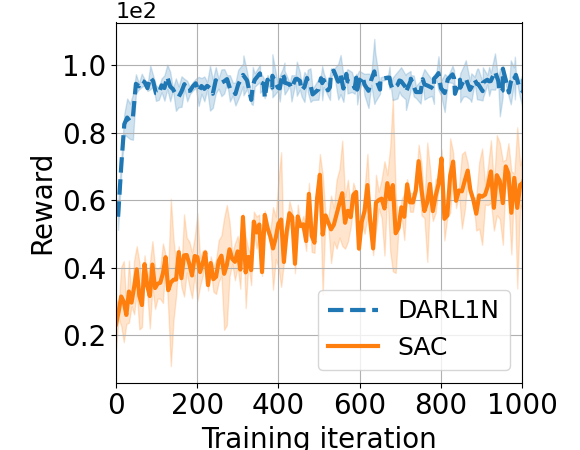}}
    \caption{Average total training reward of DARL1N and SAC in the Multi-Access Wireless Communication environment when (a) $z=2$ (b) $z=10$.}
    \label{fig:sac_mwc}
    \vspace{-0.5cm}
\end{figure}

In this environment, DARL1N again achieves good performance in terms of policy quality and training efficiency compared to the benchmark methods, as shown in Tab.~\ref{tab:adv_conv} and Fig.~\ref{fig:adv_tt}. To further evaluate the performance, we reconsider the last scenario ($M=48$) and train the good agents and adversary agents using two different methods. The trained good agents and adversarial agents then compete with each other. %
We apply the Min-Max normalization to measure the normalized total reward of agents at each side achieved in an episode. To reduce uncertainty, we generate 10 episodes and record the mean values and standard deviations.  
As shown in Fig.~\ref{fig:adv_score}(b), DARL1N achieves the best performance, and both DARL1N and EPC significantly outperform MADDPG and MFAC.

\paragraph{Multi-Access Wireless Communication}
Fig. \ref{fig:sac_mwc} shows 
that SAC achieves a higher reward than DARL1N when $z$ takes a small value. 
However, when $z$ increases, which causes an exponential growth of the state space, DARL1N achieves a much higher reward and converges much faster than SAC. 
This demonstrates that DARL1N scales better than SAC with the size of the state space.

\subsubsection{Impact of Neighbor Distance}
The parameter of neighbor distance $d$ in DARL1N determines the number of one-hop neighbors of an agent, thereby influencing both training efficiency and policy quality. To evaluate its impact, we conduct experiments using the Grassland environment, considering three good and three adversary agents. The rewards are set to $10$ for good agents collecting a grass pellet and $-100$ for colliding with adversary agents.

\begin{figure}[t]
    \centering
    {\small (a)\hspace*{-0.05em}}\subfigure{\includegraphics[width=0.445\linewidth]{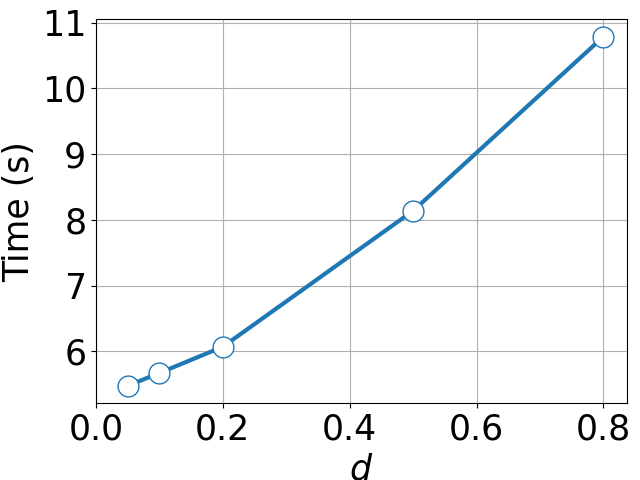}}\hspace*{-0.25em}
    {\small (b)\hspace*{-0.05em}}\subfigure{\includegraphics[width=0.455\linewidth]{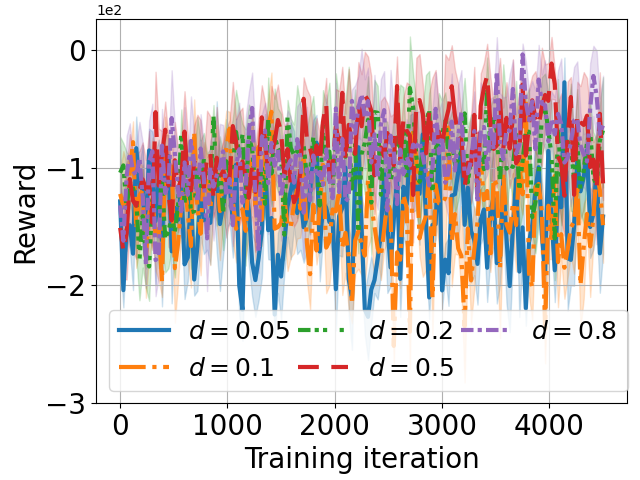}}\hspace*{-1em}
    \caption{(a) Average training time of 30 iterations and (b) average total training reward of DARL1N in the Grassland environment when $d$ increases with $M=6$.}
    \label{fig:varying_neighbor_distances}
    \vspace{-2.0em}
\end{figure}

The results shown in Fig. \ref{fig:varying_neighbor_distances} indicate that,  as the neighbor distance $d$ increases, the training reward increases while the training time arises. This stems from the increased number of one-hop neighbors each agent must consider, thereby requiring each learner to collect and process more data.
This phenomenon reveals a trade-off between training efficiency and policy quality controlled by the neighbor distance, which can be properly chosen to achieve a good balance. 

\subsection{Performance of Coded Distributed Learning Architecture}
\label{sec:coded_darl1n_exp}
In this section, we first conduct numerical studies to evaluate the performance of different assignment schemes described in Sec. \ref{sec:assignment}. We then train DARL1N over the proposed coded distributed learning architecture and conduct  experimental studies  to evaluate its performance in mitigating the effect of \revision{computing} stragglers.

\begin{table}[t]
          \caption{\revision{Comparison of different assignment schemes in terms of computation overhead, success rate, and average $V$.}}
           \label{tab:numerical_scheme}
	\resizebox{\linewidth}{!}{
 \begin{tabular}{|c|c|c|c|c|c|c|c|c|}
		\hline
		 \multicolumn{2}{|c|}{} & \multirow{2}{*}{\textbf{\makecell{Computation \\ overhead}}} & \multicolumn{3}{c|}{\textbf{Success rate}} &   \multicolumn{3}{c|}{\textbf{Average V}}  \\
    \cline{4-7}\cline{8-9}
          \multicolumn{2}{|c|}{}  &   & $\eta=0$ & $\eta=0.2$ & $\eta=0.5$ & $|\calJ|=12$ & $|\calJ|=18$ & $|\calJ|=24$\\
        \hline
        \multicolumn{2}{|c|}{\textbf{Uncoded}} & \textbf{0} & \textbf{1} & 0 & 0 & \textbf{0} &  0 & 0 \\
        \hline
        \multicolumn{2}{|c|}{\textbf{MDS}} & 23 & \textbf{1} & \textbf{1} & \textbf{1} & 124.82 &  79.75 & 73.64 \\
        \hline
         \multirow{3}{*}{\textbf{Random Sparse}} & $\xi=0.2$  & 4.5  & \textbf{1}  & 0.78 & 0.17 & 11.48 & 1.62 & -0.68\\
         \cline{2-9}
        & $\xi=0.4$  & 8.0 & \textbf{1}  & \textbf{1} & 0.98 & 16.82 &  2.93 & -1.28\\
         \cline{2-9}
        & $\xi=0.8$  & 18.3  & \textbf{1}  & \textbf{1} & \textbf{1} & 13.15 & 1.40 & -3.31\\
       \hline
        \multicolumn{2}{|c|}{\textbf{Repetition}} & 1 & \textbf{1} & 0.603 & 0 & \textbf{0} & 0 & \textbf{-8.32} \\
        \hline
         \multirow{3}{*}{\textbf{LDGM}} & $\rho=0.1$  & 0.91  & \textbf{1}  & 0.253 & 0 & 2.08 &  \textbf{-1.46} & -4.59\\
         \cline{2-9}
        & $\rho=0.3$  & 4.41  & \textbf{1}  & 0.79 & 0.11 & 8.93 & 1.05  & -3.21\\
         \cline{2-9}
        & $\rho=0.5$ & 5.5  & \textbf{1} & 0.99 & 0.41   & 13.13 & 5.13 & -0.31\\
         \hline
	\end{tabular}}
 \vspace{-1.5em}
 \end{table}

\subsubsection{Numerical Evaluation}

\revision{We conduct numerical studies to evaluate the performance of different assignment schemes in the following three aspects. 
\begin{itemize}
    \item \textit{Computation overhead}: 
 Metric \eqref{eq:computation_overhead} is applied, with the mean overhead averaged over 10 experiment runs used for the Random Sparse and LDGM schemes.
 
    \item \textit{Resilience to stragglers}: The success rate computed as follows is used. We randomly turn some learners into stragglers that fail to return any results according to Assumption~\ref{assumption:straggler}. Monte Carlo simulations are then conducted to measure the success rate, which is the ratio of training iterations in which gradients can be successfully estimated with results returned from non-stragglers.
    
    \item \textit{Impact on policy quality}: According to Theorem \ref{theorem:convergence}, the gradients estimated by Coded DARL1N 
    are unbiased but their variance depends on the assignment matrix. Therefore, we use the variance of the estimated gradients, denoted by $\mathbb{V}[\hat{\bfe}]$, to assess the impact of assignment schemes on policy quality. Specifically, we vary the number of learners whose results are used by the central controller to estimate the gradients and calculate the average value of $V: = \log(\det(\mathbb{V}[\tilde{\bfe}]))$ over $100$ Monte Carlo simulation runs.
\end{itemize}}

\revision{Tab.~\ref{tab:numerical_scheme} presents results when $M=12$ and $N=24$. The performance of the Random Sparse and LDGM schemes, characterized by parameters $\xi$ and $\rho$ respectively, is evaluated across different parameter values.
The results show that the Uncoded scheme has the lowest computation overhead, the smallest variance in most scenarios, but the poorest resilience to stragglers. In contrast, the MDS scheme exhibits the best resilience to stragglers but the largest computation overhead and variance. The Repetition scheme has the smallest variance among all schemes and the lowest computation overhead among coded schemes, though it is relatively less resilient. 
For the Random Sparse and LDGM schemes, increasing $\xi$ or $\rho$ leads to higher computation overhead and improved resilience to stragglers. However, the Random Sparse scheme generally exhibits larger variance compared to LDGM. In the following experiments, we set $\xi=0.8$ and $\rho=0.3$.}

\subsubsection{Experiments}
To understand the performance of the coded distributed learning architecture, we train DARL1N using different assignment schemes and evaluate its performance in different straggler scenarios \revision{simulated on Amazon EC2}.  

\paragraph{Experiment Settings}
We select the Food Collection environment and set the number of agents and learners in all experiments to $M=12$ and $N=24$, respectively. To evaluate the impact of stragglers, we vary the straggler probability $\eta$ in Assumption \ref{assumption:straggler}. \revision{
As Amazon EC2 computing instances are generally stable, we simulate 
stragglers by having selected compute nodes
delay returning results by $\Delta>0$ amount of time. 
Evaluations on other computing systems where stragglers are more common, such as wireless and mobile computing systems, are deferred to future work.}

\begin{figure}[t]
    \centering
    \includegraphics[width=1\linewidth]{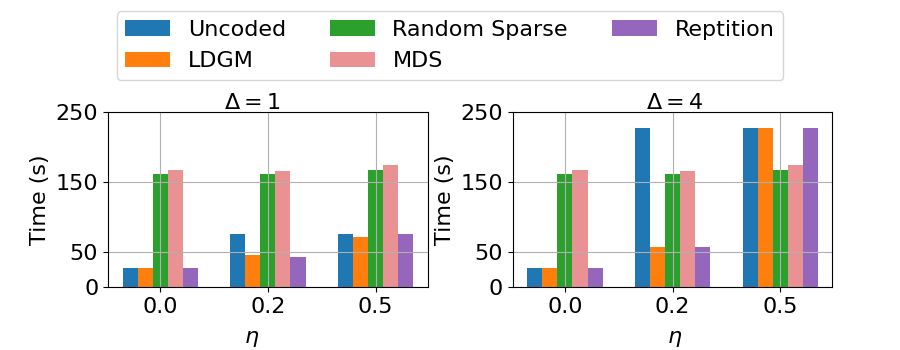}
    \vspace{-1.4em}
    \caption{Average training time of different DARL1N implementations with straggler effect $\Delta = 1$ and $\Delta = 4$, respectively.}
    \label{fig:coded_time}
    \vspace{-1.4em}
\end{figure}

\paragraph{Experiment Results}
We first evaluate the average training time of DARL1N with different assignment schemes. We vary the straggler probability $\eta$ and the straggler effect $\Delta$. The results are shown in Fig.~\ref{fig:coded_time}(a) when $\Delta=1$ and Fig.~\ref{fig:coded_time}(b) when $\Delta=4$, where the training time is measured by averaging the time for running 30 training iterations. 
We can observe that when no stragglers exist ($\eta=0$), the Uncoded scheme is the most efficient as it has zero computation overhead. The MDS and Random Sparse schemes require a much longer training time than other schemes due to the substantial computation overhead introduced by these schemes. When stragglers exist ($\eta > 0$), the performance of the Uncoded scheme degrades significantly, especially when the straggler effect is significant as shown in Fig.~\ref{fig:coded_time}(b). Compared to the Uncoded scheme, the LDGM and Repetition schemes are more resilient to stragglers, as indicated by the slower increase in training time. They are also more efficient than the MDS and Random Sparse schemes in most cases. On the contrary, the training time of MDS and Random Sparse does not grow much as $\Delta$ increases from $1$ to $4$, evidencing their high resilience to stragglers. Although they require more training time than other schemes when the straggler effect $\Delta$ or the straggler probability $\eta$ is small, they achieve higher training efficiency when $\Delta$ and/or $\eta$ are large.

To evaluate the impact of different assignment schemes on the quality of trained policies, we measure the training reward achieved by each DARL1N implementation with $\Delta=1$. Tab.~\ref{tab:coded_conv} summarizes the convergence time, convergence reward, and variance $V$, averaged over 600 training iterations, for different implementations as straggler probability $\eta$ increases. We can see that the Uncoded scheme converges fast when no stragglers exist ($\eta=0$) but its convergence speed decreases significantly when stragglers exist ($\eta>0$). The MDS and Random Sparse schemes achieve the lowest reward and slowest convergence rate, while the LDGM scheme achieves the highest reward and convergence rate in most cases, especially when the straggler probability $\eta$ is large. The Repetition scheme generally achieves good training reward performance and converges fast when the straggler probability is small. Moreover, we can also see that a larger average V generally leads to a lower reward. 

\begin{table}[t]
\caption{Convergence time, convergence reward and average V of different DARL1N implementations.}
\label{tab:coded_conv}
\centering
\resizebox{\linewidth}{!}{\begin{tabular}{|c|c|c|c|c|c|c|c|c|c|}
	\hline
	\bf \multirow{2}{*}{Schemes} & \multicolumn{3}{c|}{Convergence Time (s)}& \multicolumn{3}{c|}{Convergence Reward} &  \multicolumn{3}{c|}{Average V}\\
	\cline{2-10}
	& $\eta=0$ &$\eta=0.2$ &$\eta=0.5$ &$\eta=0$ &$\eta=0.2$ & $\eta=0.5$ & $\eta=0$ & $\eta=0.2$ &$\eta=0.5$ \\
	\hline
    Uncoded & 323 &  510 &   521& 8 &  5 & 10 & 0 & 0 & 0\\
    \hline
    MDS & 502 &  748 &  625 & -231 &  -253 &  -212 & 82.08 &  100.10 &  104.06\\
    \hline
    Random Sparse & 512 &  820 &   670& -252 & -231 &  -227 & 15.28 &  13.47 &  12.88\\
    \hline
    Repetition & 331 &   372 &    564 & 11 &  6 &  8 & \bf{-3.16}  & \bf{-4.49} & \bf{-4.15}\\
    \hline
    LDGM  & 324 & \bf{320} &  450 & 9 & \bf{12} &  \bf{14} & -0.51 & 0.14 & 0.35\\
\hline
\end{tabular}}
\vspace{-1.8em}
\end{table}


\subsubsection{Discussion}

\revision{The experiment results above suggest guidelines for selecting an appropriate assignment scheme of agents to learners. The Uncoded scheme has zero computation overhead and low variance but is the least resilient to stragglers. This makes it suited best for stable distributed systems, such as server-based setups with wired communication. The MDS scheme, while offering the highest resilience, incurs the highest computation overhead and variance, leading to poor policy quality, and making it unsuitable for distributed training using DARL1N. In unstable distributed systems, where stragglers are present, a trade-off between policy quality and resilience must be considered. If policy quality is the priority, the Repetition scheme is an excellent choice. Conversely, if resilience is more critical, the Random Sparse scheme is preferable. For a balanced approach that addresses both aspects, the LDGM scheme is a good option.}


\section{Limitations and Future Work}
\label{sec::limitation}
\revision{
In environments with local reward and transition models, DARL1N needs a suitably chosen distance metric to establish the agent neighborhoods that achieve the right balance between policy quality and ability to distribute the training efficiently. In future work, we will explore learning a neighbor distance metric that adapts to the environment, e.g., based on past episodes or contextual information, to achieve an effective balance between policy reward and training speed. Moreover, the coded distributed learning architecture for DARL1N is designed for a distributed computing system with a stable central controller in place. In future work, we will design a new coded architecture to improve resilience of central controllers such as introducing redundant central controllers using coding theory. Other issues to consider for further improvements include integration of curriculum learning similar as EPC, partially observable states, and software infrastructure to support distributed learning with low-latency communication.}


\section{Conclusion}
\label{sec:conclusion}
This paper introduced DARL1N, a scalable MARL algorithm that can be trained over a distribute computing architecture. DARL1N reduces the representation complexity of the value and policy functions of each agent in a MARL problem by disregarding the influence of other agents that are not within one hop of a proximity graph. This model enables highly efficient distributed training, in which a compute node only needs data from an agent it is training and its potential one-hop neighbors. We conducted comprehensive experiments using five MARL environments and compared DARL1N with four state-of-the-art MARL algorithms. DARL1N generates equally good or even better policies in almost all scenarios with significantly higher training efficiency than benchmark methods, especially in large-scale problem settings. To improve the resilience of DARL1N to stragglers common in distributed computing systems, we developed coding schemes that assign each agent to multiple learners. \revision{We evaluate properties of MDS, Random Sparse, Repetition, LDGM codes and provide guidelines on selecting suitable assignment schemes under different situations. }

\begin{small}
    \bibliographystyle{IEEEtran}
    \bibliography{IEEEabrv, main}
\end{small}

\appendix
\section{Appendix}

\subsection{Proof of Lemma \ref{lemma:approximation_error}}
\label{sec:proof_lemma1}
Consider the Q-value function $Q_{i}^{\bfmu}$ of agent $i$. For two different sets of non-neighbor states $\hat{\bfs}_{\calN_{i}^-}\neq \bfs_{\calN_{i}^-}$ and actions $\hat{\bfa}_{\calN_{i}^-}\neq \bfa_{\calN_{i}^-}$, we first show that:
\begin{align}
|Q_{i}^{\bfmu}(\bfs_{\calN_{i}}, \bfs_{\calN_{i}^-}, \bfa_{\calN_{i}}, \bfa_{\calN_{i}^-}) &- Q_{i}^{\bfmu}(\bfs_{\calN_{i}}, \hat{\bfs}_{\calN_{i}^-}, \bfa_{\calN_{i}}, \hat{\bfa}_{\calN_{i}^-})| \notag\\
&\leq \frac{2\bar{r}\gamma}{1-\gamma}.
\end{align}
Letting $(\bfs, \bfa)$ and $(\hat{\bfs}, \hat{\bfa})$ denote $(\bfs_{\calN_{i}}, \bfs_{\calN_{i}^-}, \bfa_{\calN_{i}}, \bfa_{\calN_{i}^-})$ and $(\bfs_{\calN_{i}}, \hat{s}_{\calN_{i}^-}, \bfa_{\calN_{i}}, \hat{a}_{\calN_{i}^-})$, respectively, we have:
\begingroup
\allowdisplaybreaks
\begin{align}
&\left|Q_{i}^{\bfmu}\left(\bfs, \bfa \right)-Q_{i}^{\bfmu}\left(\hat{\bfs}, \hat{\bfa}\right)\right|\nonumber\\
&=\biggl\lvert\mathbb{E}[\sum_{t=0}^{\infty}\gamma^{t} r_{i}\left(\bfs_{\calN_i}(t), \bfa_{\calN_i}(t)\right) \mid(\bfs(0), \bfa(0))=(\bfs, \bfa)]\nonumber\\
&-\mathbb{E}[\sum_{t=0}^{\infty}\gamma^{t} r_{i}\left(\bfs_{\calN_i}(t), \bfa_{\calN_i}(t)\right) \mid(\bfs(0), \bfa(0))=\left(\hat{\bfs}, \hat{\bfa}\right)]\biggr\rvert\nonumber\\
&\leq \sum_{t=0}^{\infty}\bigl\lvert\mathbb{E}\left[\gamma^{t} r_{i}\left(\bfs_{\calN_i}(t), \bfa_{\calN_i}(t)\right) \mid(\bfs(0), \bfa(0))=(\bfs, \bfa)\right]\nonumber\\
&-\mathbb{E}\left[\gamma^{t} r_{i}\left(\bfs_{\calN_i}(t), \bfa_{\calN_i}(t)\right) \mid(\bfs(0), \bfa(0))=\left(\hat{\bfs}, \hat{\bfa}\right)\right]\bigr\rvert\nonumber\\
&\stackrel{(\bfa)}{=} \sum_{t=1}^{\infty}\bigl\lvert\mathbb{E}\left[\gamma^{t} r_{i}\left(\bfs_{\calN_i}(t), \bfa_{\calN_i}(t)\right) \mid(\bfs(0), \bfa(0))=(\bfs, \bfa)\right]\nonumber\\
&-\mathbb{E}\left[\gamma^{t} r_{i}\left(\bfs_{\calN_i}(t), \bfa_{\calN_i}(t)\right) \mid(\bfs(0), \bfa(0))=\left(\hat{\bfs}, \hat{\bfa}\right)\right]\bigr\rvert\nonumber\\
&\leq \sum_{t=1}^{\infty}\gamma^{t}(\bigl\lvert\mathbb{E}\left[r_{i}\left(\bfs_{\calN_i}(t), \bfa_{\calN_i}(t)\right) \mid(\bfs(0), \bfa(0))=(\bfs, \bfa)\right]\bigr\rvert\nonumber\\
&+\bigl\lvert\mathbb{E}\left[r_{i}\left(\bfs_{\calN_i}(t), \bfa_{\calN_i}(t)\right) \mid(\bfs(0), \bfa(0))=\left(\hat{\bfs}, \hat{\bfa}\right)\right]\bigr\rvert)\nonumber\\
&\leq \sum_{t=1}^{\infty}2\gamma^{t}\bar{r}=\frac{2\bar{r}\gamma}{1-\gamma}
\end{align}
\endgroup
where $(\bfa)$ derives from the fact that $(\bfs_{\calN_i}, \bfa_{\calN_i})$ are part of both $(\bfs, \bfa)$ and $(\hat{\bfs}, \hat{\bfa})$. In the above equations, the expectation $\bbE$ is over state-action trajectories generated by the policy $\bfmu$ and the transition model $p$.
Then, we have:
\begin{align}
&\left|\tilde{Q}_{i}^{\bfmu}\left(\bfs_{\calN_i}, \bfa_{\calN_i} \right)-Q_{i}^{\bfmu}\left(\bfs, \bfa\right)\right|\nonumber\\
&=\biggl|\sum_{\bfs_{\calN_{i}^-}, \bfa_{\calN_{i}^-}}\!\!\omega_i(\bfs_{\calN_i}, \bfa_{\calN_i}, \bfs_{\calN_{i}^-}, \bfa_{\calN_{i}^-})Q_i^{\bfmu}(\bfs_{\calN_i}, \bfa_{\calN_i}, \bfs_{\calN_{i}^-}, \bfa_{\calN_{i}^-})\nonumber\\
&\quad-Q_i^{\bfmu}(\bfs_{\calN_i}, \bfa_{\calN_i}, \hat{\bfs}_{\calN_{i}^-}, \hat{\bfa}_{\calN_{i}^-})\biggr|\nonumber\\
&\leq \!\!\sum_{\bfs_{\calN_{i}^-}, \bfa_{\calN_{i}^-}}\!\!\omega_i(\bfs_{\calN_i}, \bfa_{\calN_i}, \bfs_{\calN_{i}^-}, \bfa_{\calN_{i}^-})\biggl|Q_i^{\bfmu}(\bfs_{\calN_i}, \bfa_{\calN_i}, \bfs_{\calN_{i}^-}, \bfa_{\calN_{i}^-})\nonumber\\
&\quad-Q_i^{\bfmu}(\bfs_{\calN_i}, \bfa_{\calN_i}, \hat{\bfs}_{\calN_{i}^-}, \hat{\bfa}_{\calN_{i}^-})\biggr| \leq\frac{2\bar{r}\gamma}{1-\gamma}.
\end{align}

\subsection{Proof of Proposition~\ref{pro:proposition1}}
\label{sec:proof_proposition_1}
If agent $j\not\in\mathcal{P}_i(t)$, then based on the definition of potential neighbors, we have $\dist(\bfs_i(t), \bfs_j(t))>d+2\epsilon$. According to the triangle inequality, 
$\dist(\bfs_i(t), \bfs_j(t+1)) + \dist(\bfs_j(t+1), \bfs_j(t))\geq \dist(\bfs_i(t), \bfs_j(t))$, and according to Assumption 1, $\dist(\bfs_j(t+1), \bfs_j(t))\leq\epsilon$. Therefore,  $\dist(\bfs_i(t), \bfs_j(t+1))>d+\epsilon$. Using the triangle inequality again, we obtain $\dist(\bfs_i(t+1), \bfs_j(t+1)) + \dist(\bfs_i(t+1), \bfs_i(t))\geq \dist(\bfs_i(t), \bfs_j(t+1))$. As $\dist(\bfs_i(t+1), \bfs_i(t))\leq\epsilon$, we have $\dist(\bfs_i(t+1), \bfs_j(t+1))> d$. Therefore, agent $j$ will not be a one-hop neighbor of agent $i$ at time $t+1$. 

\subsection{Proof of Theorem~\ref{theorem:convergence}}
\label{sec:proof_proposition_convergence}

The bias of the gradient estimator $\tilde{\bfe}$ can be calculated using \eqref{eq:grad_estimator} and \eqref{eq:recover} as follows:
\begin{align}
    \label{eq:unbiased_estimation0}
    \mathbb{E}[\tilde{\bfe}] - \bfe
     & = (\bfC_\calJ^T\bfC_\calJ)^{-1}\bfC^T_\calJ\mathbb{E}[\bfy_\calJ] - \bfe \nonumber\\
     & = (\bfC_\calJ^T\bfC_\calJ)^{-1}\bfC^T_\calJ\bfD\mathbb{E}[\bfq] - \bfe.
\end{align}
Since each learner uses the same set of parameters broadcast by the central controller for agent-environment interaction in each training iteration, the replay buffers $\calD_{j,i}$, $\forall j\in[N]$, all follow the same distribution as that of $\calD_i$. Therefore, we have $\mathbb{E}[\hat{\bfe}_{j,i}]=\bfe_i$ and $\bfD\mathbb{E}[\bfq]=\bfC_\calJ\bfe$ leading to:
\begin{equation}\label{eq:unbiased_estimation}
    \mathbb{E}[\tilde{\bfe}] - \bfe = (\bfC_\calJ^T\bfC_\calJ)^{-1}\bfC^T_\calJ\bfC_\calJ\bfe - \bfe = 0,
\end{equation}
which shows that $\tilde{\bfe}$ is an unbiased estimator.

Next, we compute the variance of the gradient estimator $\tilde{\bfe}$:
\begin{align}
    \label{eq:variance_gradient}
    \mathbb{V}[\tilde{\bfe}] &=\mathbb{V}[(\bfC_\calJ^T\bfC_\calJ)^{-1}\bfC^T_\calJ\bfy_\calJ]\\
     & = (\bfC_\calJ^T\bfC_\calJ)^{-1}\bfC^T_\calJ \bfD\mathbb{V}(\bfq)\bfD^T ((\bfC_\calJ^T\bfC_\calJ)^{-1}\bfC^T_\calJ)^T,\nonumber
\end{align}
where $\mathbb{V}[\bfq] = \diag(\mathbb{V}(\hat{\bfe}_{1,1}),\ldots,\mathbb{V}(\hat{\bfe}_{N,M}))$
and $\diag()$ creates a diagonal matrix with the $\mathbb{V}(\hat{\bfe}_{j,i})$ as diagonal element. Since $\hat{\bfe}_{j, i}, \forall i\in [M]$ are independent from each other for each $j \in [N]$.
According to \eqref{eq:variance_gradient}, we can see that the variance of the gradient estimator $\tilde{\bfe}$ is impacted by $\bfC_\calJ$, which is determined by the assignment matrix $\bfC$ as well as the learners who return their computations promptly. 

\subsection{Proof of Proposition~\ref{prop:MDS}}
\label{sec:proof_proposition_3}

The performance of the MDS code scheme will be affected only if the number of stragglers exceeds $N-M$ because $\bfC^{\text{MDS}}$ has rank $M$. If there are $W > N-M$ stragglers, the results from non-straggler nodes will be insufficient for the central controller to decode the parameter gradients and it needs to wait for results from the stragglers. Under Assumption \ref{assumption:straggler}, $W$ follows a binomial distribution with probability mass function $p(W=w)={N\choose w}(1-\eta)^{N-w}\eta^w$. Therefore, the probability that the performance will be affected by the stragglers is $\sum_{j=N-M+1}^{N}p(W=j)=\sum_{j=N-M+1}^{N} {N\choose j}(1-\eta)^{N-j}\eta^j$.

\subsection{Proof of Proposition~\ref{prop:rep}}
\label{sec:proof_proposition_5}
\revision{The assignment matrix $\bfC^\text{Repetition}$ defined in \eqref{eq:C_repetition} has $M$ linearly independent rows, with each row containing $\frac{N}{M}$ duplicate copies. For $j$-th copy of $i$-th linearly independent row, $\forall i \in [M], \forall j \in [\frac{N}{M}]$, we have a learner with index $i + (j-1)M$ needs to send $\bfy_{i + (j-1)M}$ back to the central controller. The estimated gradients can be decoded when learners with index $i + (j-1)M, \forall i \in [M]$, with any $j\in[\frac{N}{M}]$ send results back to satisfy rank($\bfC_\calJ$) = $M$. Under Assumption \ref{assumption:straggler}, for a $i\in[M]$, the probability that the learners with index $i + (j-1)M, \forall j\in[\frac{N}{M}]$ are all stragglers that fail to send results back is $\eta^{\frac{N}{M}}$. Furthermore, the probability that there is at least one non-straggler learner for each $i\in[M]$ is $(1-\eta^{\frac{N}{M}})^M$. Therefore, the probability that the performance will be affected by stragglers is then represented with $1-(1-\eta^{\frac{N}{M}})^M$.}


\begin{IEEEbiography}
[{\includegraphics[width=1in,height=1.25in,clip,keepaspectratio]{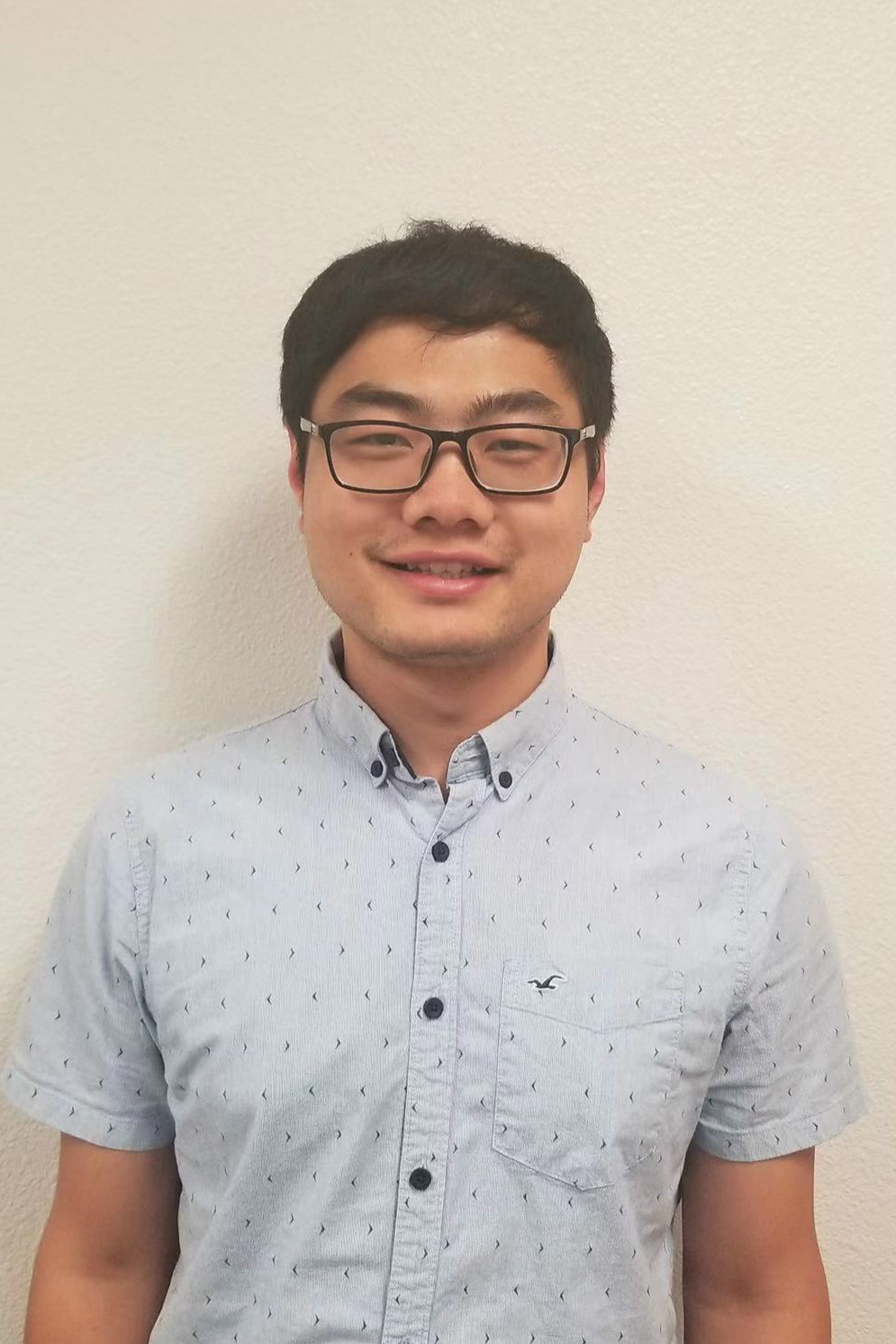}}]{Baoqian Wang} (S'20) is currently an advanced technologist in intelligent systems at The Boeing Company. He received a Ph.D. degree in Electrical and Computer Engineering from University of California San Diego and San Diego State University in 2023. He received his B.S. degree from Yangtze University, Wuhan China, in 2017, and M.S. degree in Computer Science from Texas A\&M University-Corpus Christi. His research interests include distributed computing, reinforcement learning and robotics.
\end{IEEEbiography}

\begin{IEEEbiography}[{\includegraphics[width=1in,height=1.25in,clip,keepaspectratio]{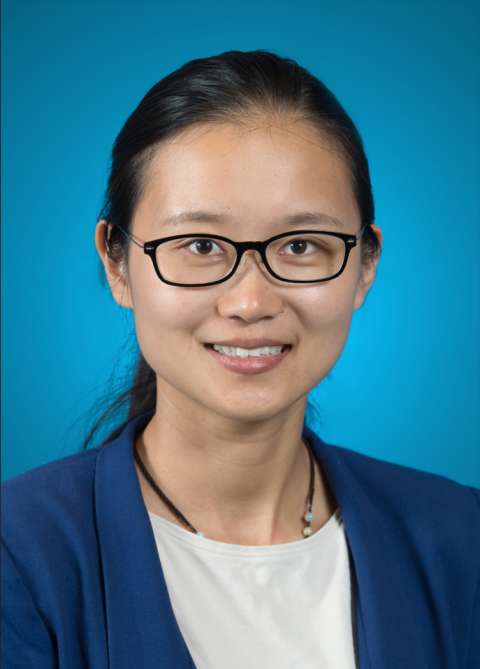}}]{Junfei Xie} (S'13-M'16-SM'21) is currently an Associate Professor in the Department of Electrical and Computer Engineering at San Diego State University. She received the B.S. degree in Electrical Engineering from University of Electronic Science and Technology of China (UESTC), Chengdu, China, in 2012. She received the M.S. degree in Electrical Engineering in 2013 and the Ph.D. degree in Computer Science and Engineering from University of North Texas, Denton, TX, in 2016. From 2016 to 2019, she was an Assistant Professor in the Department of Computing Sciences at Texas A\&M University-Corpus Christi. She is the recipient of the NSF CAREER Award. Her current research interests include large-scale dynamic system design and control, airborne networks, airborne computing, and air traffic flow management, etc.
\end{IEEEbiography}

\begin{IEEEbiography}[{\includegraphics[width=1in,height=1.25in,clip,keepaspectratio]{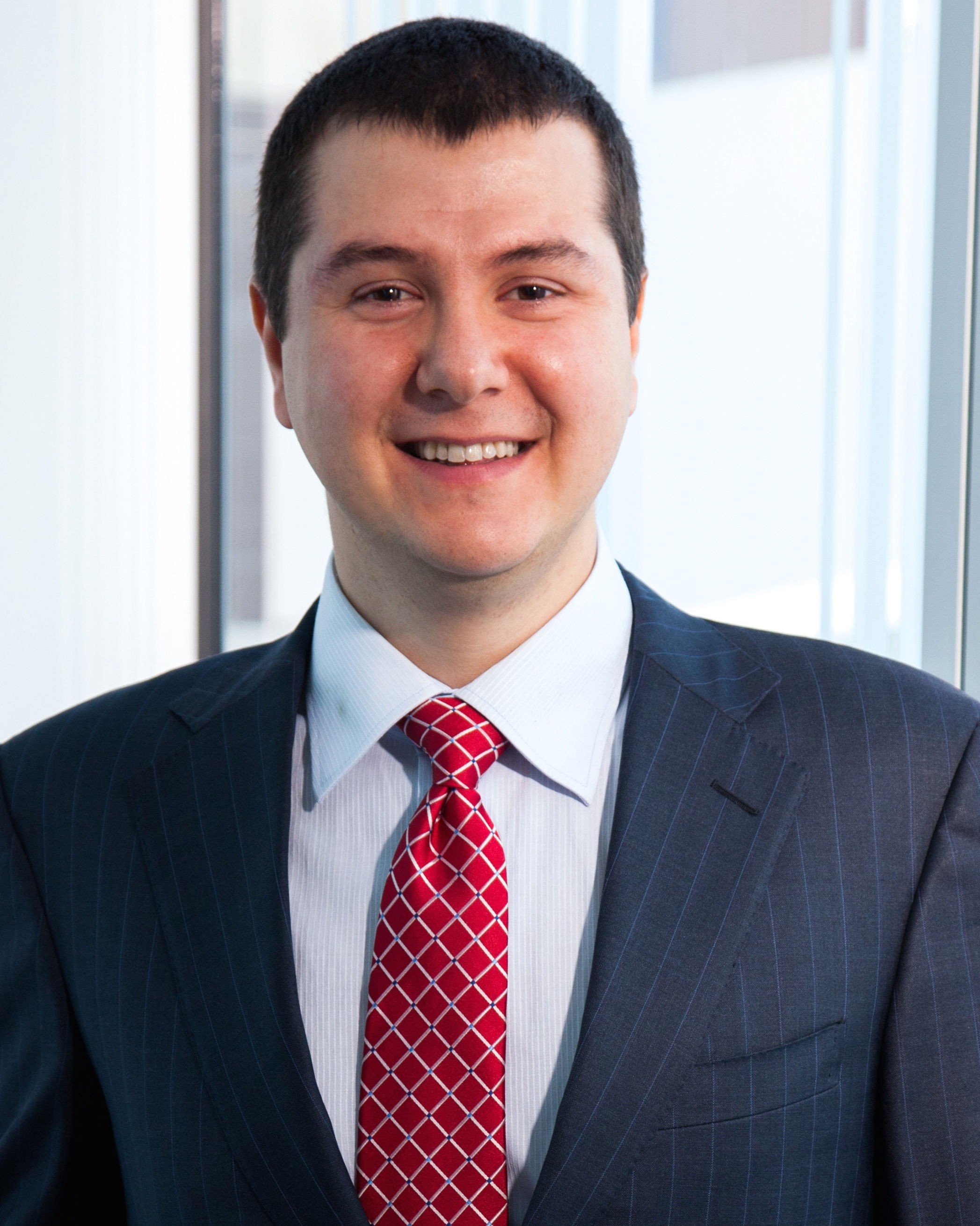}}]{Nikolay Atanasov}
(S'07-M'16-SM'23) is an Associate Professor of Electrical and Computer Engineering at the University of California San Diego, La Jolla, CA, USA. He obtained a B.S. degree in Electrical Engineering from Trinity College, Hartford, CT, USA in 2008 and M.S. and Ph.D. degrees in Electrical and Systems Engineering from the University of Pennsylvania, Philadelphia, PA, USA in 2012 and 2015, respectively. Dr. Atanasov's research focuses on robotics, control theory, and machine learning with emphasis on active perception problems for autonomous mobile robots. He works on probabilistic models for simultaneous localization and mapping (SLAM) and on optimal control and reinforcement learning algorithms for minimizing probabilistic model uncertainty. Dr. Atanasov's work has been recognized by the Joseph and Rosaline Wolf award for the best Ph.D. dissertation in Electrical and Systems Engineering at the University of Pennsylvania in 2015, the Best Conference Paper Award at the IEEE International Conference on Robotics and Automation (ICRA) in 2017, the NSF CAREER Award in 2021, and the IEEE RAS Early Academic Career Award in Robotics and Automation in 2023.
\end{IEEEbiography}
\end{document}